\documentclass{ws-rv9x6}  

\begin{document}

\setcounter{chapter}{0}

\chapter{PHASE DIAGRAM OF NEUTRAL QUARK MATTER AT MODERATE DENSITIES}

\markboth{R\"uster, Werth, Buballa, Shovkovy, and Rischke}{Phase 
diagram of neutral quark matter at moderate densities}

\author{Stefan B.\ R\"uster}
\address{Institut f\"ur Theoretische Physik,
J.W.\ Goethe-Universit\"at,\\
D-60438 Frankfurt am Main, Germany\\
E-mail: ruester@th.physik.uni-frankfurt.de}
\vspace{-1mm}

\author{Verena Werth}
\address{Institut f\"ur Kernphysik,
Technische Universit\"at Darmstadt,\\
D-64289 Darmstadt, Germany\\
E-mail: verena.werth@physik.tu-darmstadt.de}
\vspace{-1mm}

\author{Michael Buballa}
\address{Institut f\"ur Kernphysik,
Technische Universit\"at Darmstadt,\\
D-64289 Darmstadt, Germany\\
E-mail: michael.buballa@physik.tu-darmstadt.de}
\vspace{-1mm}

\author{I.~A.~Shovkovy}
\address{Frankfurt Institute for Advanced Studies,
J.W.\ Goethe-Universit\"at, \\
D-60438 Frankfurt am Main, Germany\\
E-mail: i.shovkovy@fias.uni-frankfurt.de}
\vspace{-1mm}

\author{D.~H.~Rischke}
\address{Institut f\"ur Theoretische Physik,
J.W.\ Goethe-Universit\"at, \\
D-60438 Frankfurt am Main, Germany\\
E-mail: drischke@th.physik.uni-frankfurt.de}
\vspace{-1mm}

\begin{abstract}
We discuss the phase diagram of moderately dense, locally neutral 
three-flavor quark matter using the framework of an effective 
model of quantum chromodynamics with a local interaction. The phase 
diagrams in the plane of temperature and quark chemical potential 
as well as in the plane of temperature and lepton-number chemical 
potential are discussed. 
\end{abstract}

\section{Introduction}     
\label{Introduction}

Theoretical studies suggest that baryon matter at sufficiently high 
density and sufficiently low temperature is a color superconductor. 
(For reviews on color superconductivity see, for example, 
Ref.~\refcite{reviews}.) In nature, the highest densities of matter are 
reached in central regions of compact stars. There, the density might
be as large as $10 \rho_{0}$ where $\rho_{0}\approx 0.15$ fm$^{-3}$ 
is the saturation density. It is possible that baryonic matter is 
deconfined under such conditions and, perhaps, it is 
color-superconducting. 

In compact stars, matter in the bulk is neutral with respect to the
electric and color charges. Matter should also remain in $\beta$ 
equilibrium. Taking these constraints consistently into account may 
have a strong effect on the competition between different phases 
of deconfined quark matter at large baryon 
densities.\cite{absence2sc,SRP,mei,g2SC,gCFL} 

The first attempts to obtain the phase diagram of dense, locally 
neutral three-flavor quark matter as a function of the strange 
quark mass, the quark chemical potential, and the temperature 
were made in Refs.~\refcite{SRP} and \refcite{phase-d}. This 
was done within the framework of a Nambu--Jona-Lasinio (NJL) 
model. It was shown that, at zero temperature and small values 
of the strange quark mass, the ground state of matter corresponds
to the color-flavor-locked (CFL) phase.\cite{cfl,weakCFL} At some 
critical value of the strange quark mass, this is replaced by the 
gapless CFL (gCFL) phase.\cite{gCFL} In addition, several other 
phases were found at nonzero temperature. For instance, it was 
shown that there should exist a metallic CFL (mCFL) phase, a 
so-called uSC phase,\cite{dSC} as well as the standard two-flavor 
color-superconducting (2SC) phase\cite{cs,weak} and the gapless 
2SC (g2SC) phase.\cite{g2SC} 

In Ref.~\refcite{phase-d}, the effect of the strange quark mass 
was incorporated only approximately through a shift of the chemical
potential of strange quarks, $\mu_{s} \to \mu_s - m_s^2/(2\mu)$. 
While such an approach is certainly reliable at small values of the 
strange quark mass, it becomes uncontrollable with increasing the 
mass. The phase diagram of Ref.~\refcite{phase-d} was further developed 
in Refs.~\refcite{phase-d1} and \refcite{phase-d2} where the shift-approximation in 
dealing with the strange quark was not employed any more, although 
quark masses were still treated as free parameters, rather 
than dynamically generated quantities. In Refs.~\refcite{pd-mass} and 
\refcite{pd-nu} the phase diagram of dense, locally neutral three-flavor quark
matter was further improved by treating dynamically generated quark 
masses self-consistently. Some results within this approach at zero 
and nonzero temperatures were also obtained in Refs.~\refcite{SRP},
\refcite{kyoto}, \refcite{pd-other} and \refcite{kyoto2}.

The results of Refs.~\refcite{pd-mass} and \refcite{pd-nu} are presented 
here. Only locally neutral phases are considered. This excludes, for example, 
mixed\cite{mix} and crystalline\cite{cryst} phases. Taking them into 
account requires a special treatment. We also discuss the effect 
of a nonzero neutrino (or, more precisely, lepton-number) chemical 
potential on the structure of the phase diagram.\cite{pd-nu} 
This is expected to have a potential relevance for the physics of 
protoneutron stars where neutrinos are trapped during the first 
few seconds of the stellar evolution.

The effect of neutrino trapping on color-superconducting quark matter
has been previously discussed in Ref.~\refcite{SRP}. There it was found 
that a nonzero neutrino chemical potential favors the 2SC phase and 
disfavors the CFL phase. This is not unexpected because the 
neutrino chemical potential is related to the conserved lepton 
number in the system and therefore it also favors the presence 
of (negatively) charged leptons. This helps 2SC-type pairing
because electrical neutrality in quark matter can be achieved 
without inducing a very large mismatch between the Fermi surfaces 
of up and down quarks. The CFL phase, on the other hand, is 
electrically and color neutral {\it in the absence} of charged 
leptons when $T =0$.\cite{enforced} A nonzero neutrino 
chemical potential can only spoil CFL-type pairing.

A more systematic survey of the phase diagram in the space of 
temperature, quark and lepton-number chemical potentials was
performed in Ref.~\refcite{pd-nu}. In particular, this included 
the possibility of gapless phases, which have not been taken into 
account in Ref.~\refcite{SRP}. While such phases are generally unstable at 
zero temperature,\cite{instability} this is not always the case at nonzero 
temperature.\cite{F-proc} Keeping this in mind, we shall merely localize 
the ``problematic" regions in the phase diagram, where unconventional 
pairing is unavoidable. We shall refrain, however, from speculating 
on various possibilities for the true ground state (see, e.g., 
Refs.~\refcite{mix,cryst,spin-1,meson-cond,hong,gluonic}), since these 
are still under debate.

In the application to protoneutron stars, it is of interest to 
cover a range of parameters that could provide a total lepton 
fraction in quark matter of up to about 0.4. This is the value 
of the lepton-to-baryon charge ratio in iron cores of progenitor 
stars. Because of the conservation of both lepton and baryon 
charges, this value is also close to the lepton fraction in 
protoneutron stars at early times, when the leptons did not 
have a chance to diffuse through dense matter and escape from 
the star.\cite{prakash-et-al} In the model introduced in the 
next section, almost the whole range of possibilities will be 
covered by restricting the quark chemical potential to 
$\mu\lesssim 500~\mbox{MeV}$ and the neutrino chemical 
potential to $\mu_{\nu_{e}}\lesssim 400~\mbox{MeV}$.

\section{Model}
\label{model}

Let us start by introducing the effective model of QCD used 
in the analysis. This is a three-flavor quark model with a 
local NJL-type interaction, whose Lagrangian density is given 
by 
\begin{eqnarray}
\mathcal{L} &=& \bar \psi \, ( i \not{\hspace{-2pt}\partial}
- \hat{m} \, ) \psi 
+G_S \sum_{a=0}^8 \left[ \left( \bar \psi \lambda_a \psi \right)^2 
+ \left( \bar \psi i \gamma_5 \lambda_a \psi \right)^2 \right] 
\nonumber \\
&+& G_D \sum_{\gamma,c} \left[\bar{\psi}_{\alpha}^{a} i \gamma_5
\epsilon^{\alpha \beta \gamma}
\epsilon_{abc} (\psi_C)_{\beta}^{b} \right] \left[ 
(\bar{\psi}_C)_{\rho}^{r} i \gamma_5
\epsilon^{\rho \sigma \gamma} \epsilon_{rsc} \psi_{\sigma}^{s} 
\right] 
\nonumber \\
&-& K \left\{ \det_{f}\left[ \bar \psi \left( 1 + \gamma_5 \right) \psi
\right] + \det_{f}\left[ \bar \psi \left( 1 - \gamma_5 \right) \psi
\right] \right\} \;,
\label{Lagrangian}
\end{eqnarray}
where the quark spinor field $\psi_{\alpha}^{a}$ carries color 
($a=r,g,b$) and flavor ($\alpha=u,d,s$) indices. The matrix of quark 
current masses is given by $\hat{m} = \mbox{diag}_{f}(m_u, m_d, m_s)$.
Regarding other notations, $\lambda_a$ with $a=1,\ldots,8$ are 
the Gell-Mann matrices in flavor space, and $\lambda_0\equiv 
\sqrt{2/3} \,\mathbf{1}_{f}$. The charge conjugate spinors are 
defined as follows: $\psi_C = C \bar \psi^T$ and $\bar 
\psi_C = \psi^T C$, where $\bar\psi=\psi^\dagger \gamma^0$ 
is the Dirac conjugate spinor and $C=i\gamma^2 \gamma^0$ 
is the charge conjugation matrix.

The model in Eq.~(\ref{Lagrangian}) should be viewed as an 
effective model of strongly interacting matter that captures at 
least some key features of QCD dynamics. The Lagrangian density 
contains three different interaction terms which are chosen to 
respect the symmetries of QCD. Note that we include the 't Hooft 
interaction whose strength is determined by the coupling constant 
$K$. This term breaks the $U(1)$ axial symmetry.

The term in the second line of Eq.~(\ref{Lagrangian}) describes a scalar
diquark interaction in the color antitriplet and flavor antitriplet
channel. For symmetry reasons there should also be a pseudoscalar 
diquark interaction with the same coupling constant. This term 
would be important to describe Goldstone boson condensation in the 
CFL phase.\cite{goldstones} 

We use the following set of model parameters:\cite{RKH}
\begin{subequations}
\label{model-parameters}
\begin{eqnarray}
&& m_{u}=m_{d} = 5.5~\mathrm{MeV} , \quad 
m_s = 140.7~\mathrm{MeV} , \\
&& G_S \Lambda^2 = 1.835  , \qquad 
K \Lambda^5 = 12.36 , \quad 
\Lambda = 602.3~\mathrm{MeV} .
\label{Lambda} 
\end{eqnarray}
\end{subequations}
After fixing the masses of the up and down quarks at equal values, 
$m_{u,d}=5.5~\mbox{MeV}$, the other four parameters are chosen to 
reproduce the following four observables of vacuum QCD:
$m_{\pi}=135.0~\mbox{MeV}$, $m_{K}=497.7~\mbox{MeV}$,
$m_{\eta^\prime}=957.8~\mbox{MeV}$, and $f_{\pi}=92.4~\mbox{MeV}$.\cite{RKH} 
This parameter set gives $m_{\eta}=514.8~\mbox{MeV}$.

In Ref.~\refcite{RKH}, the diquark coupling $G_D$ was not fixed by the 
fit of the meson spectrum in vacuum. In general, it is expected to be 
of the same order as the quark-antiquark coupling $G_S$. Here, 
we choose the coupling strength with $G_D=\frac34 G_S$ which follows
from the Fierz identity.

The grand partition function, up to an irrelevant normalization 
constant, is given by 
\begin{equation}
\label{Z}
\mathcal{Z} \equiv e^{-\Omega V/T}
= \int \mathcal{D} \bar\psi \mathcal{D} \psi \, e^{i
\int_X \left( \mathcal{L} + \bar\psi \hat{\mu} \gamma^0 \psi
\right) } \; ,
\end{equation}
where $\Omega$ is the thermodynamic potential density, $V$ is the 
volume of the three-space, and $\hat\mu$ is a diagonal matrix of 
quark chemical potentials. In chemical equilibrium (which provides 
$\beta$ equilibrium as a special case), the nontrivial components 
of this matrix are extracted from the following relation:
\begin{equation}
\mu_{ab}^{\alpha\beta} = \left(
  \mu \delta^{\alpha\beta} 
+ \mu_Q Q_{f}^{\alpha\beta} \right)\delta_{ab} 
+ \left[ \mu_3 \left(T_3\right)_{ab} 
+ \mu_8 \left(T_8\right)_{ab} \right] \delta^{\alpha\beta} \; .
\label{mu-f-i}
\end{equation}
Here $\mu$ is the quark chemical potential (by definition, 
$\mu = \mu_{B}/3$ where $\mu_{B}$ is the baryon chemical 
potential), $\mu_Q$ is the chemical potential of electric charge, 
while $\mu_3$ and $\mu_8$ are color chemical potentials associated 
with two mutually commuting color charges of the $SU(3)_c$ gauge 
group, cf. Ref.~\refcite{BuSh}. 
The explicit form of the electric charge matrix is 
        $Q_{f}=\mbox{diag}_{f}(\frac23,-\frac13,-\frac13)$,
and the explicit form of the color charge matrices is
        $T_3=\mbox{diag}_c(\frac12,-\frac12,0)$ and 
$\sqrt{3}T_8=\mbox{diag}_c(\frac12,\frac12,-1)$.

In order to calculate the mean-field thermodynamic potential 
at temperature $T$, we first linearize the interaction in the 
presence of the diquark condensates $\Delta_{c} \sim 
(\bar{\psi}_C)_{\alpha}^{a} i \gamma_5 \epsilon^{\alpha \beta c} 
\epsilon_{a b c} \psi_{\beta}^{b}$ (no sum over $c$) and the 
quark-antiquark condensates $\sigma_\alpha \sim \bar \psi_\alpha^a 
\psi_\alpha^a$ (no sum over $\alpha$). Then, integrating out the 
quark fields and neglecting the fluctuations of composite order 
parameters, we arrive at the following expression for the 
thermodynamic potential:
\begin{eqnarray}
\Omega &=& \Omega_{L} +
\frac{1}{4 G_D} \sum_{c=1}^{3} \left| \Delta_c \right|^2
+2 G_S \sum_{\alpha=1}^{3} \sigma_\alpha^2 \nonumber\\
&-& 4 K \sigma_u \sigma_d \sigma_s
-\frac{T}{2V} \sum_K \ln \det \frac{S^{-1}}{T} \; ,
\label{Omega}
\end{eqnarray}
where we also added the contribution of leptons, $\Omega_{L}$, 
which will be specified later. 

We should note that we have restricted ourselves to field 
contractions corresponding to the Hartree approximation. 
In a more complete treatment, among others, 
the 't Hooft interaction term gives also rise 
to mixed contributions containing both diquark and quark-antiquark 
condensates, i.e., $\propto \sum_{\alpha=1}^{3} \sigma_\alpha 
|\Delta_\alpha|^2$.\cite{Steiner} In this study, 
we neglect such terms for simplicity. While their presence may change 
the results quantitatively, one does not expect them to modify the 
qualitative structure of the phase diagram. 

In Eq.~(\ref{Omega}), $S^{-1}$ is the inverse full quark propagator 
in the Nambu-Gorkov representation, 
\begin{equation}
S^{-1} = 
\left(
\begin{array}{cc}
[ G_0^+ ]^{-1} & \Phi^- \\
\Phi^+ & [ G_0^- ]^{-1}
\end{array}
\right) \; ,
\label{off-d}
\end{equation}
with the diagonal elements being the inverse Dirac 
propagators of quarks and of charge-conjugate quarks, 
\begin{equation}
[ G_0^\pm ]^{-1} = \gamma^\mu K_\mu 
\pm \hat \mu \gamma_0 - \hat{M} \; ,
\end{equation}
where $K^\mu = (k_0, \mathbf{k})$ denotes the four-momentum of the 
quark. At nonzero temperature, we use the Matsubara imaginary time 
formalism. Therefore, the energy $k_0$ is replaced with $-i \omega_n$ 
where $\omega_n\equiv (2n+1)\pi T$ are the fermionic Matsubara 
frequencies. Accordingly, the sum over $K$ in Eq.~(\ref{Omega}) 
should be interpreted as a sum over integer $n$ and an integral 
over the three-momentum $\mathbf{k}$.

The constituent quark mass matrix $\hat{M} = \mbox{diag}_{f}(M_u,M_d,M_s)$ 
is defined by 
\begin{equation}
M_\alpha = m_\alpha - 4 G_S \sigma_\alpha 
+ 2 K \sigma_\beta \sigma_\gamma \; ,
\label{Mi}
\end{equation}
where $\sigma_\alpha $ are the quark-antiquark condensates, and
the set of indices $(\alpha, \beta, \gamma)$ is a permutation of 
$(u,d,s)$.

The off-diagonal components of the inverse propagator (\ref{off-d}) 
are the so-called gap matrices given in terms of three diquark
condensates. The color-flavor structure of these matrices is 
given by
\begin{equation}
\left(\Phi^-\right)^{\alpha\beta}_{ab} 
= -\sum_c \epsilon^{\alpha\beta c}\,\epsilon_{abc}
   \,\Delta_c\,\gamma_5~,
\label{Phim}
\end{equation}
and $\Phi^+ = \gamma^0 (\Phi^-)^\dagger \gamma^0$. Here, as before, 
$a$ and $b$ refer to the color components and $\alpha$ and $\beta$ 
refer to the flavor components. Hence, the gap parameters $\Delta_1$, 
$\Delta_2$, and $\Delta_3$ correspond to the down-strange, the up-strange 
and the up-down diquark condensates, respectively. All three of them 
originate from the color-antitriplet, flavor-antitriplet diquark 
pairing channel. For simplicity, the color and flavor symmetric 
condensates are neglected in this study. They were shown to be 
small and not crucial for the qualitative understanding of the 
phase diagram.\cite{phase-d}

The determinant of the inverse quark propagator can be decomposed 
as follows:\cite{pd-mass}
\begin{equation}
   \det \frac{S^{-1}}{T}
=  \prod_{i=1}^{18} 
   \left( \frac{ \omega_n^2 + \epsilon_i ^2 }{T^2}
   \right)^2 \; ,
\label{det-S}
\end{equation}
where $\epsilon_i$ are eighteen independent positive energy eigenvalues. 
The Matsubara summation in Eq.~(\ref{Omega}) can then be done analytically 
by employing the relation\cite{Kapusta}
\begin{equation}
\sum_n \ln \left( \frac{ \omega_n^2 + \epsilon_i^2}{T^2} \right) 
= \frac{| \epsilon_i|}{T} 
+ 2 \ln \left( 1 + e^{- \frac{| \epsilon_i |}{T}} \right) \; .
\end{equation}
Then, we arrive at the following mean-field expression for the 
pressure:
\begin{eqnarray}
p \equiv -\Omega &=& \frac{1}{2 \pi^2} \sum_{i=1}^{18} \int_0^\Lambda d k \, k^2
\left[ |\epsilon_i| + 2 T \ln \left( 1 + e^{-
\frac{|\epsilon_i|}{T}} \right) \right] \nonumber \\ 
&+& 4 K \sigma_u \sigma_d \sigma_s
-\frac{1}{4 G_D} \sum_{c=1}^{3} \left| \Delta_c \right|^2
-2 G_S \sum_{\alpha=1}^{3} \sigma_\alpha^2
 \nonumber \\ 
&+& \frac{T}{\pi^2} \sum_{l=e,\mu} \sum_{\epsilon=\pm}
\int_0^\infty d k \, k^2
\ln \left( 1 + e^{-\frac{E_l-\epsilon\mu_l}{T}}\right) \nonumber \\ 
&+& \frac{1}{24\pi^2} \sum_{l=e,\mu} \left( \mu_{\nu_l}^4 +
2 \pi^2 \mu_{\nu_l}^2 T^2 + \frac{7}{15} \pi^4
T^4 \right)\; ,
\label{pressure}
\end{eqnarray}
where the contributions of electrons and muons with masses 
$m_e \approx 0.511$ MeV and $m_\mu \approx 105.66$ MeV, as well 
as the contributions of neutrinos were included. Note that muons
may exist in matter in $\beta$ equilibrium and, therefore, they are 
included in the model for consistency. However, being about 200 times 
heavier than electrons, they do not play a big role in the analysis.

\section{Phase diagram in absence of neutrino trapping}

In this section, we consider the case without neutrino trapping in 
quark matter. This is expected to be a good approximation for matter 
inside a neutron star after the short deleptonization period is over. 

The expression for the pressure in Eq.~(\ref{pressure}) has a physical 
meaning only when the chiral and color-superconducting order parameters,
$\sigma_\alpha$ and $\Delta_c$, satisfy the following set of six gap equations:
\begin{eqnarray}
&& \frac{\partial p}{\partial \sigma_\alpha} = 0 \;  , \quad
\frac{\partial p}{\partial \Delta_c} = 0 \; .
\label{gapeqns}
\end{eqnarray}
To enforce the conditions of local charge neutrality in dense matter,
we also require three other equations to be satisfied,
\begin{eqnarray}
&& n_Q \equiv \frac{ \partial p }{\partial \mu_Q} = 0 \; , \quad 
n_3 \equiv \frac{ \partial p }{\partial \mu_3} = 0 \; , \quad 
n_8 \equiv \frac{ \partial p }{\partial \mu_8} = 0 \; .
\label{neutrality}
\end{eqnarray}
These fix the values of the three corresponding chemical potentials,
$\mu_Q$, $\mu_3$ and $\mu_8$. After these are fixed, only the quark 
chemical potential $\mu$ is left as a free parameter.

In order to obtain the phase diagram, one has to find the ground 
state of matter for each given set of the parameters in the model.
In the case of locally neutral matter, there are two parameters
that should be specified: temperature $T$ and quark chemical potential 
$\mu$. After these are fixed, one has to compare the values of the 
pressure in all competing neutral phases of quark matter. The ground 
state corresponds to the phase with the highest pressure. 

By using standard numerical recipes, it is 
not extremely difficult to find a solution to the given set of 
nine nonlinear equations. Complications arise, however, due to 
the fact that often the solution is not unique. 
The existence of different solutions to the same set of equations, 
(\ref{gapeqns}) and (\ref{neutrality}), reflects the physical 
fact that there could exist several competing neutral phases with 
different physical properties. Among these phases, all but one 
are unstable or metastable. In order to take this into account, 
one should look for the solutions of the following 8 types:

\begin{enumerate}

\item Normal quark (NQ) phase: $\Delta_1=\Delta_2=\Delta_3=0$;

\item 2SC phase: $\Delta_1=\Delta_2=0$,  and $\Delta_3\neq 0$;

\item 2SC${us}$ phase: $\Delta_1=\Delta_3=0$,  and $\Delta_2\neq 0$;

\item 2SC${ds}$ phase: $\Delta_2=\Delta_3=0$,  and $\Delta_1\neq 0$;

\item uSC phase:  $\Delta_2\neq 0$, $\Delta_3\neq 0$, and $\Delta_1= 0$;

\item dSC phase: $\Delta_1\neq 0$, $\Delta_3\neq 0$, and $\Delta_2= 0$;

\item sSC phase: $\Delta_1\neq 0$, $\Delta_2\neq 0$, and $\Delta_3= 0$;

\item CFL phase: $\Delta_1\neq0$, $\Delta_2\neq 0$, $\Delta_3\neq 0$.

\end{enumerate}

Then, we calculate the values of the pressure in all nonequivalent 
phases, and determine the ground state as the phase with the highest 
pressure. After this is done, we study additionally the spectrum of 
low-energy quasiparticles in search for the existence of gapless modes. 
This allows us to refine the specific nature of the ground state. 

In the above definition of the eight phases in terms of $\Delta_c$,
we have ignored the quark-antiquark condensates $\sigma_\alpha$.
In fact, in the chiral limit ($m_\alpha= 0$), the quantities 
$\sigma_\alpha$ are good order parameters and we could define 
additional sub-phases characterized by nonvanishing values of 
one or more $\sigma_\alpha$. With the model parameters at hand, 
however, chiral symmetry is broken explicitly by the nonzero current 
quark masses, and the values of $\sigma_\alpha$ never vanish. 
Hence, in a strict sense it is impossible to define any new phases 
in terms of $\sigma_\alpha$. 

Of course, this does not exclude the possibility of discontinuous 
changes in $\sigma_\alpha$ at some line in the plane of temperature 
and quark chemical potential, thereby constituting a first-order 
phase transition line. It is generally expected that the ``would-be'' 
chiral phase transition remains first-order at low temperatures, 
even for nonzero quark masses. Above some critical temperature, 
however, this line could end in a critical endpoint and there is 
only a smooth crossover at higher temperatures. This picture 
emerges from NJL-model studies, both, without\cite{cr-pt-NJL} 
and with\cite{BO} diquark pairing. 

\begin{figure}[ht]
  \begin{center}
    \includegraphics[width=0.75\textwidth]{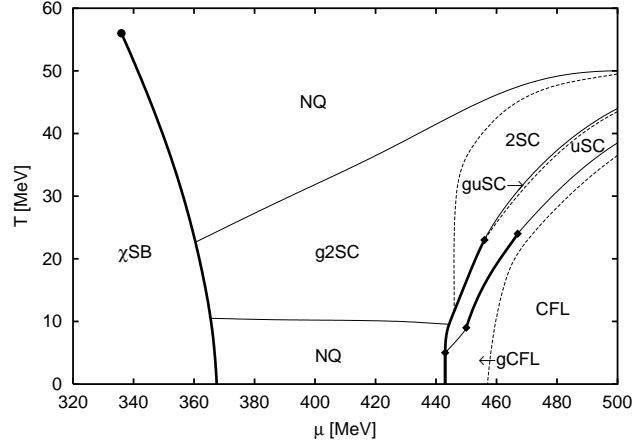}
    \caption{The phase diagram of neutral quark matter in model
     (\ref{Lagrangian}).
     First-order phase boundaries are indicated by bold solid lines,
     whereas the thin solid lines mark second-order phase boundaries
     between two phases which differ by one or more nonzero diquark
     condensates. The dashed lines indicate the (dis-)appearance of 
     gapless modes in different phases, and they do not correspond 
     to phase transitions.}
    \label{phasediagram}
  \end{center}
\end{figure}

\begin{figure}[ht]
  \begin{center}
    \includegraphics[width=0.95\textwidth]{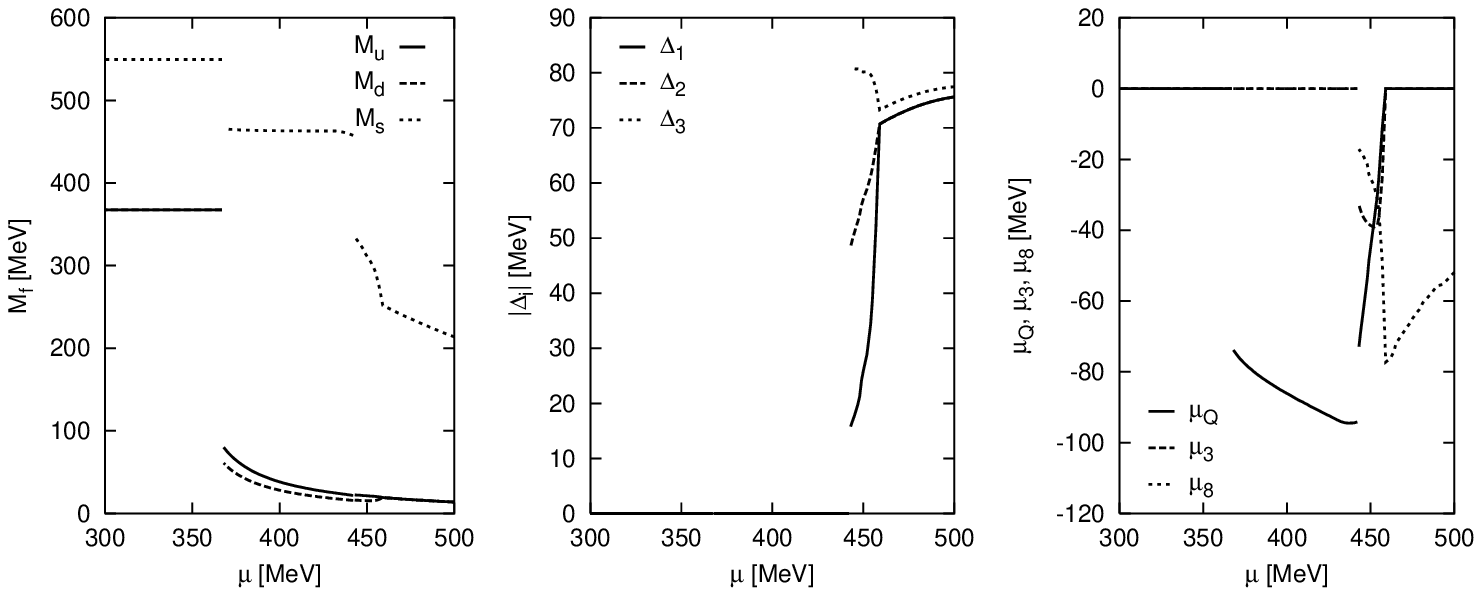}\\
    \includegraphics[width=0.95\textwidth]{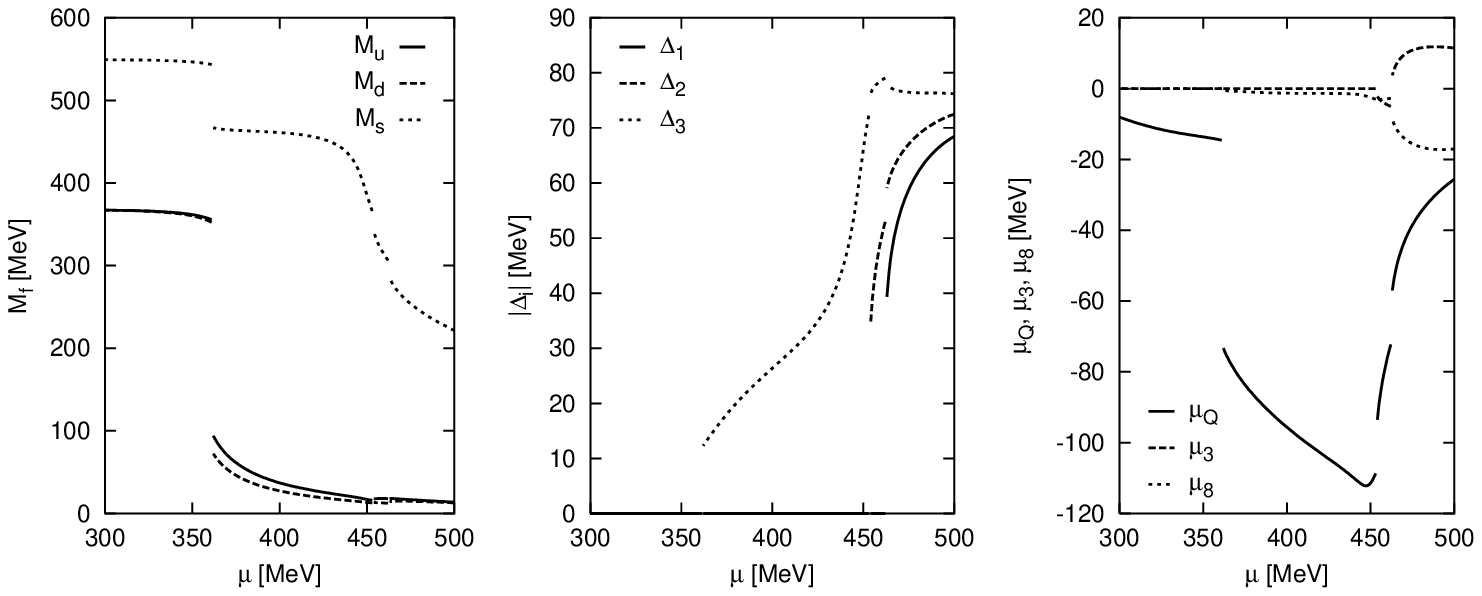}\\
    \includegraphics[width=0.95\textwidth]{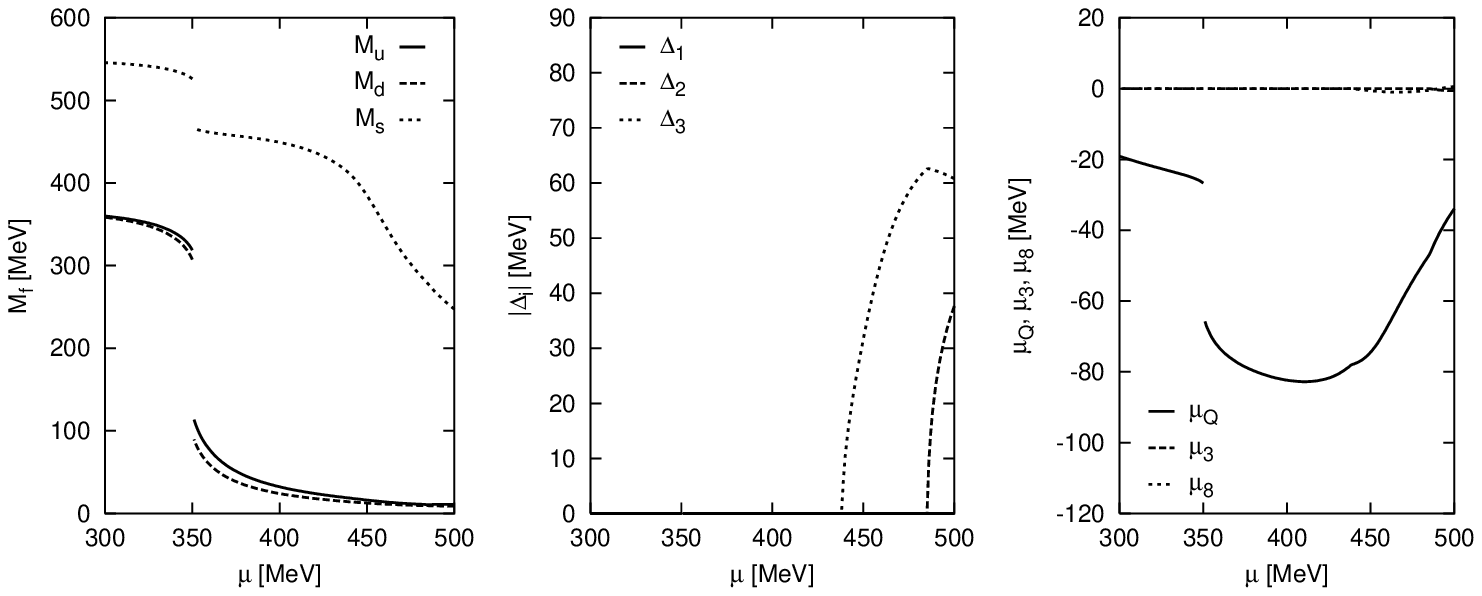}
    \caption{Dependence of the quark masses, of the gap parameters, 
     and of the electric and color charge chemical potentials on the 
     quark chemical 
     potential at a fixed temperature, $T=0~\mbox{MeV}$ (three upper 
     panels), $T=20~\mbox{MeV}$ (three middle panels), and $T=40~\mbox{MeV}$ 
     (three lower panels).}
    \label{plot0-20-40}
  \end{center}
\end{figure}

The numerical results for neutral quark matter are summarized in 
Fig.~\ref{phasediagram}. This shows the phase diagram in the plane 
of temperature and quark chemical potential, obtained in the 
mean-field approximation in model (\ref{Lagrangian}). The 
corresponding dynamical quark masses, 
gap parameters, and three charge chemical potentials are displayed 
in Fig.~\ref{plot0-20-40}. All quantities are plotted as functions 
of $\mu$ for three different values of the temperature: 
$T=0,~20,~40$~MeV. 


In the region of small quark chemical potentials 
and low temperatures, the phase diagram is dominated by the normal 
phase in which the approximate chiral symmetry is broken, and in 
which quarks have relatively large constituent masses. This is 
denoted by $\chi$SB in Fig.~\ref{phasediagram}. With increasing 
the temperature, this phase changes smoothly into the NQ phase 
in which quark masses are relatively small. Because of explicit 
breaking of the chiral symmetry in the model at hand, there 
is no need for a phase transition between the two regimes. 

However, as pointed out above, the symmetry argument does not 
exclude the possibility of an ``accidental'' (first-order) chiral 
phase transition. As expected, at lower temperatures we find a line 
of first-order chiral phase transitions. It is located within a 
relatively narrow window of the quark chemical potentials 
($336~\mbox{MeV} \lesssim \mu \lesssim 368~\mbox{MeV}$) which are 
of the order of the vacuum values of the light-quark constituent 
masses. (For the parameters used in the calculation one obtains 
$M_u=M_d=367.7\mbox{~MeV}$ and $M_s=549.5\mbox{~MeV}$ in 
vacuum.\cite{RKH}) At this critical line, the quark chiral condensates, 
as well as the quark constituent masses, change discontinuously. 
With increasing temperature, the size of the discontinuity decreases, 
and the line terminates at the endpoint located at $(T_{\rm cr},
\mu_{\rm cr})\approx (56,336)~\mbox{MeV}$, see Fig.~\ref{phasediagram}. 

The location of the critical endpoint is consistent with other 
mean-field studies of NJL models with similar sets of 
parameters.\cite{cr-pt-NJL,BO} This agreement does not need to be exact 
because, in contrast to the studies in Refs.~\refcite{cr-pt-NJL,BO}, 
here we imposed the condition of electric charge neutrality in quark 
matter. (Note that the color neutrality is satisfied automatically in 
the normal phase.) One may argue, however, that the additional 
constraint of neutrality is unlikely to play a big role in the 
vicinity of the endpoint. 

It is appropriate to mention here that the location of the critical 
endpoint might be affected very much by fluctuations of the composite 
chiral fields. These are not included in the mean-field studies of
the NJL model. In fact, this is probably the main reason for their 
inability to pin down the location of the critical endpoint consistent,
for example, with lattice calculations.\cite{lattice} (It is 
fair to mention that the current lattice calculations are not very 
reliable at nonzero $\mu$ either.) Therefore, the predictions of this 
study, as well as of those in Refs.~\refcite{cr-pt-NJL} and \refcite{BO}, regarding 
the critical endpoint cannot be considered as very reliable. 

When the quark chemical potential exceeds some critical value and the 
temperature is not too large, a Cooper instability with respect to 
diquark condensation should develop in the system. Without enforcing 
neutrality, i.e., if the chemical potentials of up and down quarks are 
equal, this happens immediately after the chiral phase transition 
when the density becomes nonzero.\cite{BO} In the present model, 
this is not the case at low temperatures.

\begin{figure}[ht]
  \begin{center}
    \includegraphics[width=0.75\textwidth]{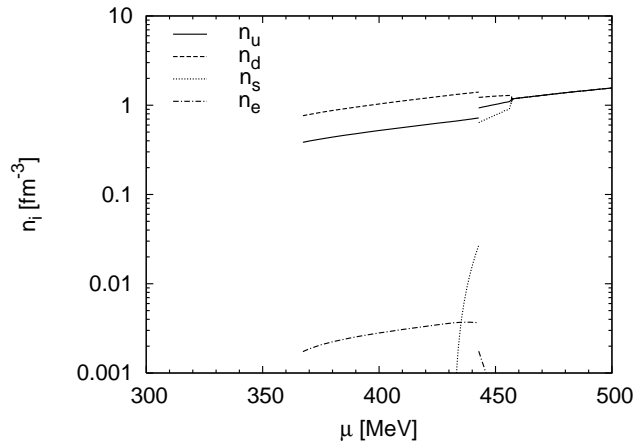}
    \caption{The dependence of the number densities of quarks and
     electrons on the quark chemical potential at $T=0~\mbox{MeV}$. 
     Note that the densities of all three quark flavors coincide 
     above $\mu=457~\mbox{MeV}$. The density of muons vanishes for
     all values of $\mu$.}
    \label{densT0log}
  \end{center}
\end{figure}

In order to understand this, one should inspect the various quantities 
at $T=0$ which are displayed in the upper three panels of 
Fig.~\ref{plot0-20-40}. At the chiral phase boundary, the up and 
down quark masses become relatively small, whereas the strange 
quark mass experiences only a moderate drop of about $84$~MeV 
induced by the 't Hooft interaction. This is not sufficient to 
populate any strange quark states at the given chemical potential, 
and the system mainly consists of up and down quarks together with 
a small fraction of electrons, see Fig.~\ref{densT0log}. The electric 
charge chemical potential which is needed to maintain neutrality in 
this regime is between about $-73$ and $-94$~MeV. It turns out that 
the resulting splitting of the up and down quark Fermi momenta is too 
large for the given diquark coupling strength to enable diquark pairing 
and the system stays in the normal phase. 

At $\mu\approx 432~\mbox{MeV}$, the chemical potential felt by the 
strange quarks, $\mu - \mu_Q/3$, reaches the strange quark mass and 
the density of strange quarks becomes nonzero. At first, this density 
is too small to play a sizable role in neutralizing matter, or in 
enabling strange-nonstrange cross-flavor diquark pairing, see 
Fig.~\ref{densT0log}. The NQ 
phase becomes metastable against the gapless CFL (gCFL) phase at 
$\mu_{\rm gCFL} \approx 443~\mbox{MeV}$. This is the point of a 
first-order phase transition. It is marked by a drop of the 
strange quark mass by about $121$~MeV. As a consequence, strange 
quarks become more abundant and pairing gets easier. Yet, in the 
gCFL phase, the strange quark mass is still relatively large, and 
the standard BCS pairing between strange and light (i.e., up and 
down) quarks is not possible. In contrast to the regular CFL phase, 
the gCFL phase requires a nonzero density of electrons to stay 
electrically neutral. At $T=0$, therefore, one could use the value 
of the electron density as a formal order parameter that distinguishes 
these two phases.\cite{gCFL}

With increasing the chemical potential further (still at $T=0$), 
the strange quark mass decreases and the cross-flavor Cooper pairing 
gets stronger. Thus, the gCFL phase eventually turns into the regular 
CFL phase at $\mu_{\rm CFL} \approx 457~\mbox{MeV}$. The electron 
density goes to zero at this point, as it should. This is indicated 
by the vanishing value of $\mu_Q$ in the CFL phase, see the upper 
right panel in Fig.~\ref{plot0-20-40}. We remind that the CFL 
phase is neutral because of having equal number densities of all 
three quark flavors, $n_u=n_d=n_s$, see Fig.~\ref{densT0log}. 
This equality is enforced by the pairing 
mechanism, and this is true even when the quark masses are not 
exactly equal.\cite{enforced} 

Let us mention here that the same NJL model at zero temperature was 
studied previously in Ref.~\refcite{SRP}. The results of Ref.~\refcite{SRP}
agree with those presented here only when the quark chemical potential 
is larger than the critical value for the transition to the CFL phase
at $457\mbox{~MeV}$. The appearance of the gCFL phase for $443\lesssim 
\mu \lesssim 457\mbox{~MeV}$ was not recognized in Ref.~\refcite{SRP}, 
however. Instead, it was suggested that there exists a narrow (about 
$12\mbox{~MeV}$ wide) window of values of the quark chemical potential 
around $\mu\approx 450\mbox{~MeV}$ in which the 2SC phase is the ground 
state. By examining the same region, we find that the 2SC phase does 
not appear there.

\begin{figure}[ht]
  \begin{center}
    \includegraphics[width=0.75\textwidth]{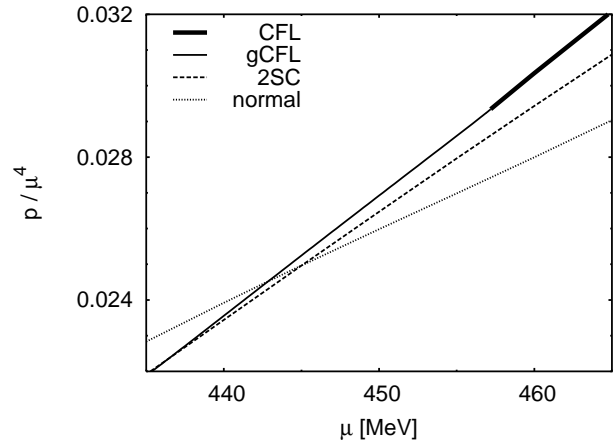}\\
    \caption{Pressure divided by $\mu^4$ for different neutral solutions
             of the gap equations at $T = 0$ as functions of the
             quark chemical potential $\mu$: regular CFL 
             (bold solid line), gapless CFL (thin solid line),
             2SC (dashed line), normal (dotted line).}
    \label{pmu0}
  \end{center}
\end{figure}

This is illustrated in Fig.~\ref{pmu0} where the pressure of three 
different solutions is displayed. Had we ignored the gCFL solution 
(thin solid line), the 2SC solution (dashed line) would indeed be the 
most favored one in the interval between $\mu \approx 445$~MeV and 
$\mu \approx 457$~MeV. After including the gCFL phase in the analysis, 
this is no longer the case.

Now let us turn to the case of nonzero temperature. One might suggest 
that this should be analogous to the zero-temperature case, except that
Cooper pairing is somewhat suppressed by thermal effects. In contrast 
to this naive expectation, the thermal distributions of quasiparticles
together with the local neutrality conditions open qualitatively new 
possibilities that were absent at $T=0$. As in the case of the two-flavor 
model of Ref.~\refcite{g2SC}, a moderate thermal smearing of mismatched 
Fermi surfaces could increase the probability of creating zero-momentum 
Cooper pairs without running into a conflict with Pauli blocking. This 
leads to the appearance of several color-superconducting phases that 
could not exist at zero temperatures.

With increasing the temperature, the first qualitatively new feature 
in the phase diagram appears when $5 \lesssim T \lesssim 
10~\mbox{MeV}$. In this temperature interval, the NQ phase is 
replaced by the uSC phase when the quark chemical potential 
exceeds the critical value of about $444\mbox{~MeV}$. The 
corresponding transition is a first-order phase transition, see 
Fig.~\ref{phasediagram}. Increasing the chemical potential 
further by several MeV, the uSC phase is then replaced by the gCFL 
phase, and the gCFL phase later turns gradually into the (m)CFL 
phase. (In this study, we do not distinguish between the CFL phase 
and the mCFL phase.\cite{phase-d}) Note that, in the model at hand, 
the transition between the uSC and the gCFL phase is of second order 
in the following two temperature intervals: $5 \lesssim T \lesssim 
9~\mbox{MeV}$ and $T \gtrsim 24~\mbox{MeV}$. On the other hand, it 
is a first-order transition when $9 \lesssim  T \lesssim 24~\mbox{MeV}$. 
Leaving aside its unusual appearance, this is likely to be an ``accidental" 
property in the model for a given set of parameters.

The transition from the gCFL to the CFL phase is a smooth crossover at 
all $T\neq 0$.\cite{phase-d,phase-d1} The reason is that the electron 
density is not a good order parameter that could be used to distinguish 
the gCFL from the CFL phase when the temperature is nonzero. This is 
also confirmed by the numerical results for the electric charge chemical 
potential $\mu_Q$ in Fig.~\ref{plot0-20-40}. While at zero temperature 
the value of $\mu_Q$ vanishes identically in the CFL phase, this is not 
the case at nonzero temperatures. 

Another new feature in the phase diagram appears when the temperature 
is above about $11~\mbox{MeV}$. In this case, with increasing the quark
chemical potential, the Cooper instability happens immediately after 
the $\chi$SB phase. The corresponding critical value of the quark 
chemical potential is rather low, about $365~\mbox{MeV}$. The first 
color-superconducting phase is the gapless 2SC (g2SC) phase.\cite{g2SC} 
This phase is replaced with the 2SC phase in a crossover transition 
only when $\mu\gtrsim 445~\mbox{MeV}$. The 2SC is then followed by the 
gapless uSC (guSC) phase, by the uSC phase, by the gCFL phase 
and, eventually, by the CFL phase (see Fig.~\ref{phasediagram}). 

In the NJL model at hand, determined by the parameters in 
Eq.~(\ref{model-parameters}), we do not find the dSC phase 
as the ground state anywhere in the phase diagram. This is 
similar to the conclusion of Refs.~\refcite{phase-d} and 
\refcite{phase-d2}, but differs from that of Refs.~\refcite{dSC} 
and \refcite{phase-d1}. This should not be surprising because, 
as was noted earlier,\cite{phase-d2,kyoto2} the appearance of the dSC 
phase is rather sensitive to a specific choice of parameters 
in the NJL model.

The phase diagram in Fig.~\ref{phasediagram} has a very specific 
ordering of quark phases. One might ask if this ordering is robust
against the modification of the parameters of the model at hand. 
We can argue that some features are indeed quite robust, while others 
are not.\cite{pd-mass}

It should be clear that the appearance of color-superconducting
phases under the stress of neutrality constraints is very sensitive 
to the strength of diquark coupling. In the case of two-flavor quark 
matter, this was demonstrated very clearly in Ref.~\refcite{g2SC} at zero
as well as at nonzero temperatures. The same statement remains true in
three-flavor quark matter.\cite{pd-mass,kyoto2} 

\section{Phase diagram in presence of neutrino trapping}

In the case of neutrino trapping, the chemical potentials of 
individual quark and lepton species can be expressed in terms of six
chemical potentials according to their content of conserved charges.
For the quarks, which carry quark number, color and electric charge,
the chemical potentials were introduced in Eq.~(\ref{mu-f-i}).
For the neutrinos, which carry only lepton number, the chemical 
potentials are
\begin{equation}
   \mu_{\nu_e} = \mu_{L_e}~, \qquad
   \mu_{\nu_\mu} = \mu_{L_\mu}~.
\end{equation}
while for the electrons and muons, carrying both lepton 
number and electric charge, the chemical potentials read
\begin{equation}
   \mu_e  = \mu_{L_e} - \mu_Q~, \qquad
   \mu_\mu  = \mu_{L_\mu} - \mu_Q~.
\end{equation}
In order to obtain the phase diagram, one has to determine 
the ground state of matter for each given set of the parameters. 
In the case of locally neutral matter with trapped neutrinos, 
there are four parameters that should be specified: the 
temperature $T$, the quark chemical potential $\mu$ as well as 
the two lepton family chemical potentials $\mu_{L_e}$ and 
$\mu_{L_\mu}$. After these are fixed, the values of the pressure 
in all competing neutral phases of quark matter should be compared. 
The phase with the largest pressure is the ground state. 
For our purposes, it is sufficient to take the vanishing muon 
lepton-number chemical potential, i.e., $\mu_{L_\mu}=0$. This 
is expected to be a good approximation for matter inside a 
protoneutron star.

\subsection{General effect of neutrino trapping}
\label{simple}

As mentioned in the Introduction, neutrino trapping favors 
the 2SC phase and disfavors the CFL phase.\cite{SRP} This is 
a consequence of the modified $\beta$-equilibrium condition 
in the system. In this section, we would like to emphasize that 
this is a model-independent effect. In order to understand the 
physics behind it, it is instructive to start our consideration 
from a very simple toy model. Many of its qualitative features 
are also observed in a self-consistent numerical analysis of the 
NJL model. 

Let us first assume that strange quarks are very heavy and
consider a gas of non-interacting massless up and down
quarks in the normal phase at $T=0$. As required by $\beta$ 
equilibrium, electrons and electron neutrinos are also 
present in the system. (Note that in this section we neglect 
muons and muon neutrinos for simplicity.) 

In the absence of Cooper pairing, the densities of quarks and 
leptons are given by
\begin{equation}
  n_{u,d} \;=\; \frac{\mu_{u,d}^3}{\pi^2}~, \quad
  n_e \;=\; \frac{\mu_e^3}{3\pi^2}~, \quad
  n_{\nu_e} \;=\; \frac{\mu_{\nu_e}^3}{6\pi^2}~. 
\end{equation}
Expressing the chemical potentials through $\mu$, $\mu_Q$ and
$\mu_{L_e}$, and imposing electric charge neutrality, one arrives 
at the following relation:
\begin{equation}
    2(1 +\frac{2}{3}y)^3 \,
-\, (1 -\frac{1}{3}y)^3 \,-\,
     (x - y)^3  \;=\; 0~,
\label{toy2}
\end{equation}
where we have introduced the chemical potential ratios 
$x = \mu_{L_e}/\mu$ and $y = \mu_Q/\mu$. The above cubic 
equation can be solved for $y$ (electric chemical
potential) at any given $x$ (lepton-number chemical
potential). The result can be used to calculate the ratio
of quark chemical potentials, $\mu_d/\mu_u = (3-y)/(3+2y)$.

\begin{figure}[ht]
  \begin{center}
    \includegraphics[width=0.75\textwidth]{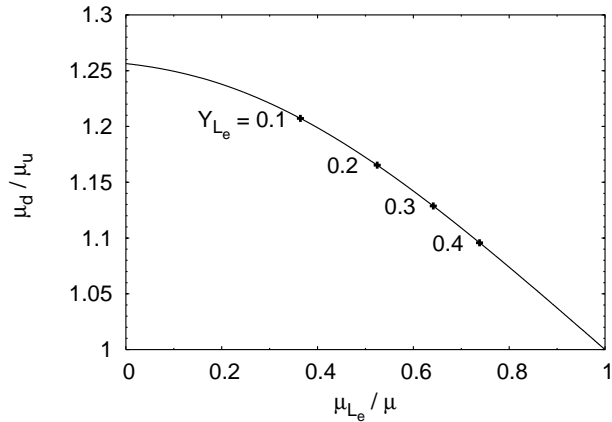}
    \caption{Ratio of down and up quark chemical potentials 
     as a function of $\mu_{L_e}/\mu$ in the toy model. 
     The crosses mark the solutions at 
     several values of the lepton fraction.}
    \label{toy}
  \end{center}
\end{figure}

The ratio $\mu_d/\mu_u$ as a function of $\mu_{L_e}/\mu$ is 
shown in Fig.~\ref{toy}. At vanishing $\mu_{L_e}$, one finds 
$y \approx -0.219$ and, thus, $\mu_d/\mu_u \approx 1.256$ 
(note that this value is very close to $2^{1/3}\approx 1.260$). 
This result corresponds to the following ratios of the number 
densities in the system: $n_u/n_d\approx 0.504$ and $n_e/n_d
\approx 0.003$, reflecting that the density of electrons is 
tiny and the charge of the up quarks has to be balanced by 
approximately twice as many down quarks.

At $\mu_{L_e}=\mu$, on the other hand, the real solution to 
Eq.~(\ref{toy2}) is $y=0$, i.e., the up and down Fermi momenta 
become equal. This can be seen most easily if one inverts the 
problem and solves Eq.~(\ref{toy2}) for $x$ at given $y$. When 
$y=0$ one finds $x=1$, meaning that $\mu_d=\mu_u$ and, in turn, 
suggesting that pairing between up and down quarks is unobstructed 
at $\mu_{L_e}=\mu$. This is in contrast to the case of vanishing 
$\mu_{L_e}$, when the two Fermi surfaces are split by about 25\%, 
and pairing is difficult. 

It is appropriate to mention that many features of the above 
considerations would not change much even when Cooper pairing 
is taken into account. The reason is that the corresponding 
corrections to the quark densities are parametrically suppressed 
by a factor of order $(\Delta/\mu)^2$. 

In order to estimate the magnitude of the effect in the case
of dense matter in protoneutron stars, we indicate several 
typical values of the lepton fractions $Y_{L_e}$ in Fig.~\ref{toy}. 
As mentioned earlier, $Y_{L_e}$ is expected to be of order $0.4$ 
right after the collapse of the iron core of a progenitor star. 
According to Fig.~\ref{toy}, this corresponds to $\mu_d/\mu_u 
\approx 1.1$, i.e., while the splitting between the up and down Fermi 
surfaces does not disappear completely, it gets reduced considerably
compared to its value in the absence of trapped neutrinos. This reduction 
substantially facilitates the cross-flavor pairing of up and down quarks. 
The effect is gradually washed out during about a dozen of seconds of 
the deleptonization period when the value of $Y_{L_e}$ decreases to zero. 

The toy model is easily modified to the opposite extreme of 
three massless quark flavors. Basically, this corresponds to 
replacing Eq.~(\ref{toy2}) by
\begin{equation}
    2(1 +\frac{2}{3}y)^3 \,-\, 2(1 -\frac{1}{3}y)^3 \,-\,
     (x - y)^3  \;=\; 0~.
\label{toy3} 
\end{equation}
In the absence of neutrino trapping, $x = 0$, the only real solution 
to this equation is $y=0$, indicating that the chemical potentials 
(which also coincide with the Fermi momenta) of up, down, and strange
quarks are equal. This reflects the fact that the system with equal 
densities of up, down, and strange quarks is neutral by itself, without 
electrons. With increasing $x\propto\mu_{L_e}$, the solution requires 
a nonzero $y\propto \mu_Q$, suggesting that up-down and up-strange 
pairing becomes more difficult. To see this more clearly, we can go 
one step further in the analysis of the toy model.

Let us assume that the quarks are paired in a regular, i.e., fully 
gapped, CFL phase at $T=0$. Then, as shown in Ref.~\refcite{enforced},
the quark part of the matter is automatically electrically neutral. 
Hence, if we want to keep the whole system electrically and color 
neutral, there must be no electrons. Obviously, this is easily 
realized without trapped neutrinos by setting $\mu_Q$ equal to zero. 
At non-vanishing $\mu_{L_e}$ the situation is more complicated. The 
quark part is still neutral by itself and therefore no electrons are 
admitted. Hence, the electron chemical potential $\mu_e = \mu_{L_e} 
- \mu_Q$ must vanish, and consequently $\mu_Q$ should be nonzero and 
equal to $\mu_{L_e}$. It is natural to ask what should be the values of
the color chemical potentials $\mu_3$ and $\mu_8$ in the CFL phase when 
$\mu_{L_e}\neq 0$. 

In order to analyze the stress on the CFL phase due to nonzero 
$\mu_{L_e}$, we follow the same approach as in Refs.~\refcite{absence2sc} 
and \refcite{gCFL}. In this analytical consideration, 
we also account for the effect of the strange quark mass simply by 
shifting the strange quark chemical potential by $-M_s^2/(2\mu)$. 
In our notation, pairing of CFL-type requires the following 
``common'' values of the Fermi momenta of paired quarks:
\begin{subequations}
\begin{eqnarray}
p_{F,(ru,gd,bs)}^{\rm common} &=& \mu -\frac{M_s^2}{6\mu},\\
p_{F,(rd,gu)}^{\rm common} &=& 
    \mu + \frac{\mu_{Q}}{6} + \frac{\mu_{8}}{2\sqrt{3}},\\
p_{F,(rs,bu)}^{\rm common} &=& 
    \mu +\frac{\mu_{Q}}{6} + \frac{\mu_{3}}{4}
   - \frac{\mu_{8}}{4\sqrt{3}} - \frac{M_s^2}{4\mu},\\
p_{F,(gs,bd)}^{\rm common} &=& 
    \mu -\frac{\mu_{Q}}{3} - \frac{\mu_{3}}{4}
   - \frac{\mu_{8}}{4\sqrt{3}} - \frac{M_s^2}{4\mu}.
\end{eqnarray}
\end{subequations}
These are used to calculate the pressure in the toy model,
\begin{eqnarray}
p^{\rm (toy)} &=& \frac{1}{\pi^2} \sum_{a=1}^{3}\sum_{\alpha=1}^{3} 
\int_{0}^{p_{F,a\alpha}^{\rm common}}\left(\mu_a^{\alpha}-p\right)
p^2 dp  \nonumber\\ 
&&+3\frac{\mu^2\Delta^2}{\pi^2}
+\frac{(\mu_{L_e}-\mu_Q)^4}{12\pi^2}+\frac{\mu_{L_e}^4}{24\pi^2}.
\end{eqnarray}
By making use of this expression, one easily derives the neutrality 
conditions as in Eq.~(\ref{neutrality}). In order to solve them, it 
is useful to note that
\begin{equation}
n_{Q}-n_{3}-\frac{1}{\sqrt{3}}n_{8} = \frac{(\mu_Q-\mu_{L_e})^3}{3\pi^2}.
\end{equation}
Thus, it becomes obvious that charge neutrality requires 
$\mu_Q=\mu_{L_e}$. The other useful observation is that the 
expression for $n_{3}$ is proportional to $\mu_3+\mu_Q$. So,
it is vanishing if (and only if) $\mu_3=-\mu_Q$, which means 
that $\mu_3=-\mu_{L_e}$. Finally, one can check that the third 
neutrality condition $n_{8}=0$ requires
\begin{equation}
\mu_{8} = -\frac{\mu_{L_e}}{\sqrt{3}}-\frac{M_{s}^{2}}{\sqrt{3}\mu}.
\end{equation}
The results for the charge chemical potentials $\mu_Q$, $\mu_3$,
and $\mu_{8}$ imply the following magnitude of stress on pairing
in the CFL phase:
\begin{subequations}
\begin{eqnarray}
\delta\mu_{(rd,gu)} &=& \frac{\mu_{g}^{u}-\mu_{r}^{d}}{2} 
                     = \mu_{L_e},\\
\delta\mu_{(rs,bu)} &=& \frac{\mu_{b}^{u}-\mu_{r}^{s}}{2}  
                     = \mu_{L_e}+\frac{M_s^2}{2\mu},\\
\delta\mu_{(gs,bd)} &=& \frac{\mu_{b}^{d}-\mu_{g}^{s}}{2}  
                     = \frac{M_s^2}{2\mu}.
\end{eqnarray}
\label{mismatch}
\end{subequations}
Note that there is no mismatch between the values of the chemical 
potentials of the other three quarks, $\mu_{r}^{u}=\mu_{g}^{d}=
\mu_{b}^{s}=\mu-M_s^2/(6\mu)$. 

{From} Eq.~(\ref{mismatch}) we see that the largest mismatch occurs 
in the $(rs,bu)$ pair (for positive $\mu_{L_e}$). The CFL phase 
can withstand the stress only if the value of $\delta\mu_{(rs,bu)}$ 
is less than $\Delta_2$. A larger mismatch should drive a transition 
to a gapless phase exactly as in Refs.~\refcite{g2SC} and \refcite{gCFL}. Thus,
the critical value of the lepton-number chemical potential is
\begin{equation}
\mu_{L_e}^{\rm (cr)} \approx \Delta_2-\frac{M_s^2}{2\mu}.
\label{mu_L^cr}
\end{equation}
When $\mu_{L_e}>\mu_{L_e}^{\rm (cr)}$, the CFL phase turns into 
the gCFL$^\prime$ phase, which is a variant of the gCFL 
phase.\cite{gCFL} By definition, the gapless mode with a linear 
dispersion relation in the gCFL$^\prime$ phase is $rs$--$bu$ 
instead of $gs$--$bd$ as in the standard gCFL phase. 
(Let us remind that the mode $a\alpha$--$b\beta$ is defined 
by its dispersion relation which interpolates between the 
dispersion relations of hole-type excitations of $a\alpha$-quark 
at small momenta, $k\ll \mu_{a}^{\alpha}$, and particle-type 
excitations of $b\beta$-quark at large momenta, 
$k\gg \mu_{b}^{\beta}$.)

In order to see what this means for the physics of protoneutron stars,
we should again try to relate the value of $\mu_{L_e}$ to the lepton 
fraction. As we have seen, there are no electrons in the (regular)
CFL phase at $T=0$. Therefore, the entire lepton number is carried 
by neutrinos. For the baryon density we may neglect the pairing 
effects to first approximation and employ the ideal-gas relations.  
This yields
\begin{equation}
Y_{L_e} \approx \frac{1}{6} \left(\frac{\mu_{L_e}}{\mu}\right)^3~.
\end{equation}
Inserting typical numbers, $\mu \gtrsim 400$~MeV and $\mu_{L_e} 
\lesssim\Delta\lesssim 100$~MeV, one finds $Y_{L_e}\lesssim 10^{-3}$. 
Thus, there is practically no chance to find a sizable 
amount of leptons in the CFL phase. The constraint gets relaxed 
slightly at nonzero temperatures and/or in the gCFL phase, but 
the lepton fraction remains rather small even then (our numerical 
results indicate that, in general, $Y_{L_e} \lesssim 0.05$ in 
the CFL phase).

\subsection{Three-dimensional phase diagram}

The simple toy-model considerations in the previous subsection give
a qualitative understanding of the effect of neutrino trapping on 
the mismatch of the quark Fermi momenta and, thus, on the pairing 
properties of two- and three-flavor quark matter. Now, we turn to 
a more detailed numerical analysis of the phase diagram in the NJL 
model defined in Sec.~\ref{model}. 

\begin{figure}[ht]
  \begin{center}
    \includegraphics[width=0.95\textwidth]{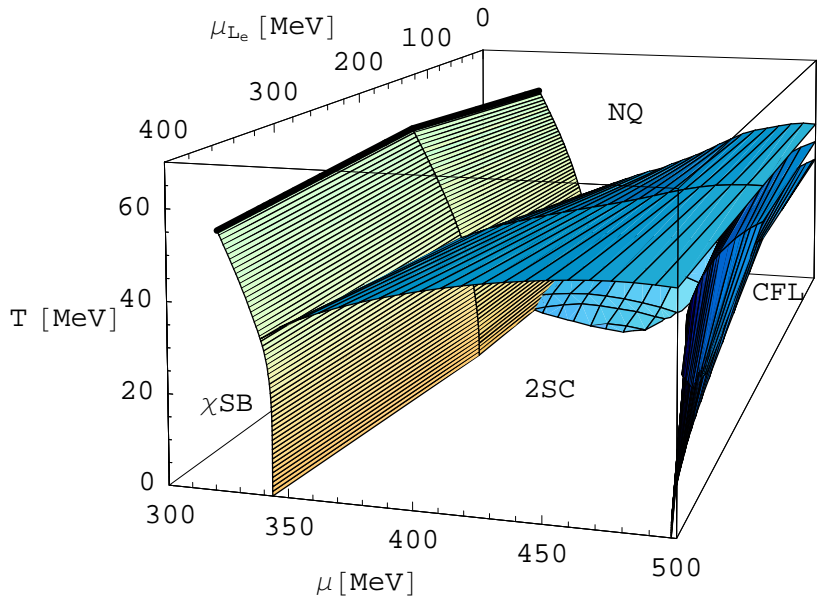}
    \caption{General structure of the phase diagram of neutral dense 
     quark matter in the three-dimensional space spanned by the quark 
     chemical potential $\mu$, the lepton-number chemical potential $\mu_{L_e}$,
     and the temperature $T$.}
    \label{phase3d}
  \end{center}
\end{figure}

The general features of the phase diagram in the three-dimensional 
space, spanned by the quark chemical potential $\mu$, the lepton-number 
chemical potential $\mu_{L_e}$, and the temperature $T$, are 
depicted in Fig.~\ref{phase3d}. Because of a rather complicated 
structure of the diagram, only the four main phases ($\chi$SB, 
NQ, 2SC and CFL) are shown explicitly. Although it is not 
labeled, a thin slice of a fifth phase, the uSC phase, squeezed 
in between the 2SC and CFL phases, can also be seen. While lacking 
detailed information, the phase diagram in Fig.~\ref{phase3d} gives 
a clear overall picture. Among other things, one sees, for example, 
that the CFL phase becomes strongly disfavored with increasing 
$\mu_{L_e}$ and gets gradually replaced by the 2SC phase.

In order to discuss the structure of the phase diagram in more detail
we proceed by showing several two-dimensional slices of it. These are 
obtained by keeping one of the chemical potentials, $\mu$ or $\mu_{L_e}$, 
fixed and varying the other two parameters. 

\subsection{$T$--$\mu$ phase diagram}

\begin{figure}[ht]
  \begin{center}
    \includegraphics[width=0.75\textwidth]{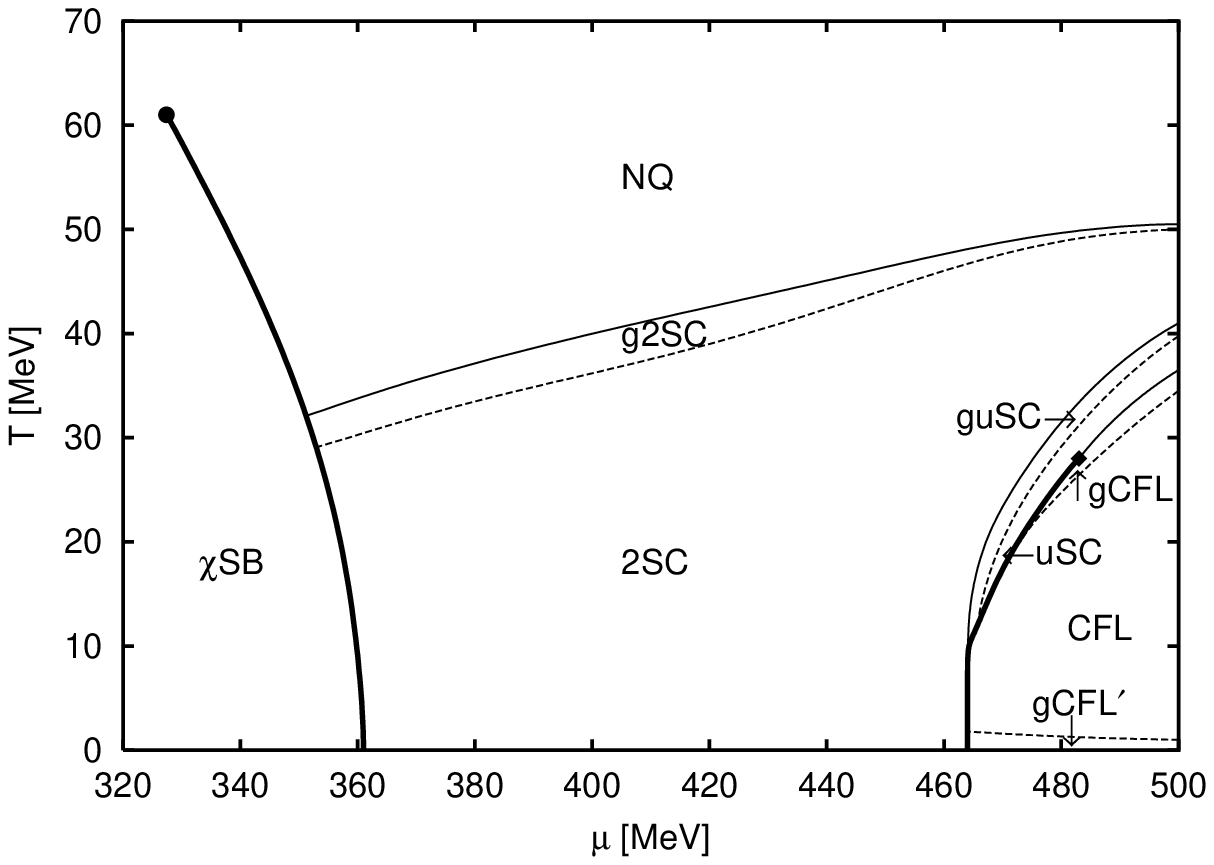}\\
    \includegraphics[width=0.75\textwidth]{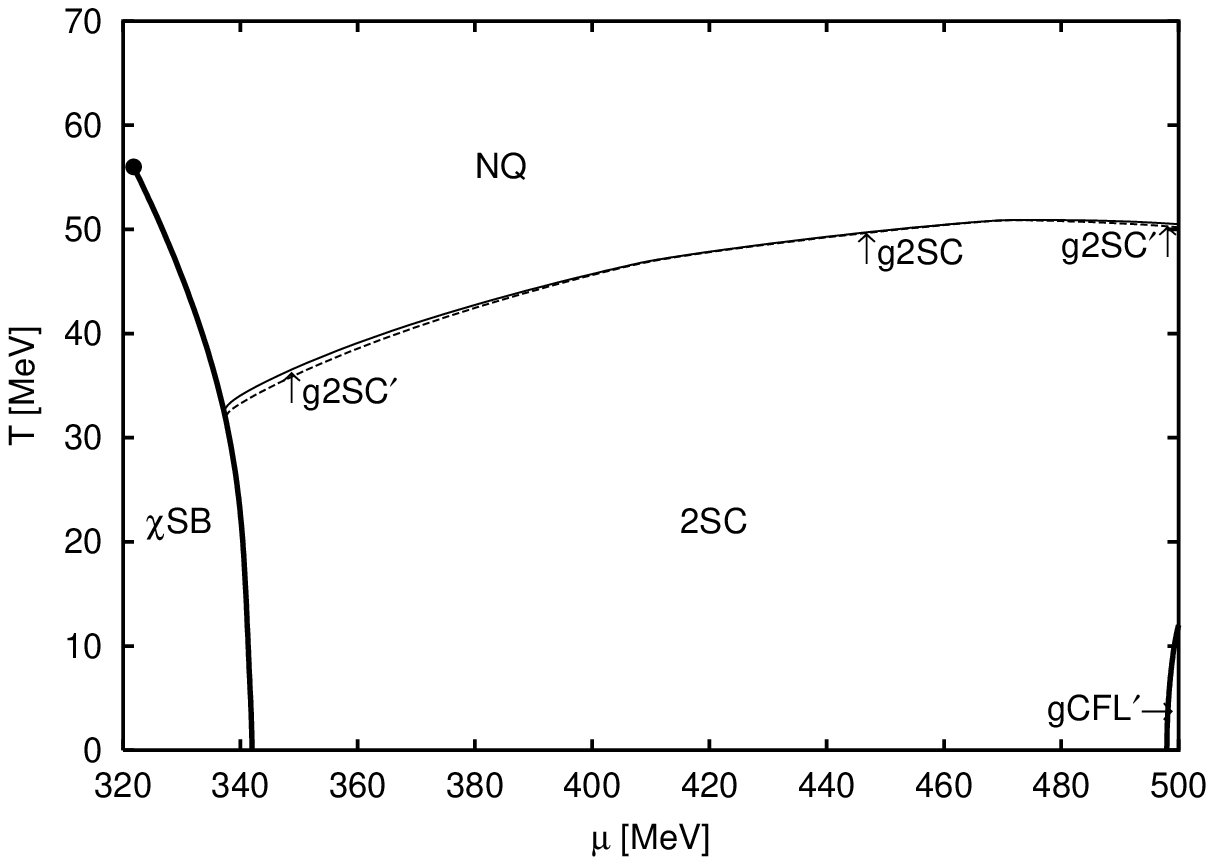}
    \caption{The phase diagrams of neutral quark matter at fixed
     lepton-number chemical potentials $\mu_{L_e}=200$~MeV 
     (upper panel), and $\mu_{L_e}=400$~MeV (lower panel), cf. 
     Fig.~\ref{phasediagram} in the absence of neutrino trapping.}
    \label{phasediagram_mu-T}
  \end{center}
\end{figure}

We begin with presenting the phase diagrams at two fixed values of the 
lepton-number chemical potential, $\mu_{L_e}=200$~MeV (upper panel) 
and $\mu_{L_e}=400$~MeV (lower panel), in Fig.~\ref{phasediagram_mu-T}. 
The general effects of neutrino trapping can be understood by analyzing 
the similarities and differences between the diagrams with and without 
neutrino trapping. 

Here it is appropriate to note that a schematic version of the 
$T$--$\mu$ phase diagram at $\mu_{L_e}=200$~MeV was first presented 
in Ref.~\refcite{SRP}, see the right panel of Fig.~4 there. If one 
ignores the complications due to the presence of the uSC phase and 
various gapless phases, the results of Ref.~\refcite{SRP} are in 
qualitative agreement with the results presented here. 

\begin{figure}[ht]
  \begin{center}
    \includegraphics[width=0.95\textwidth]{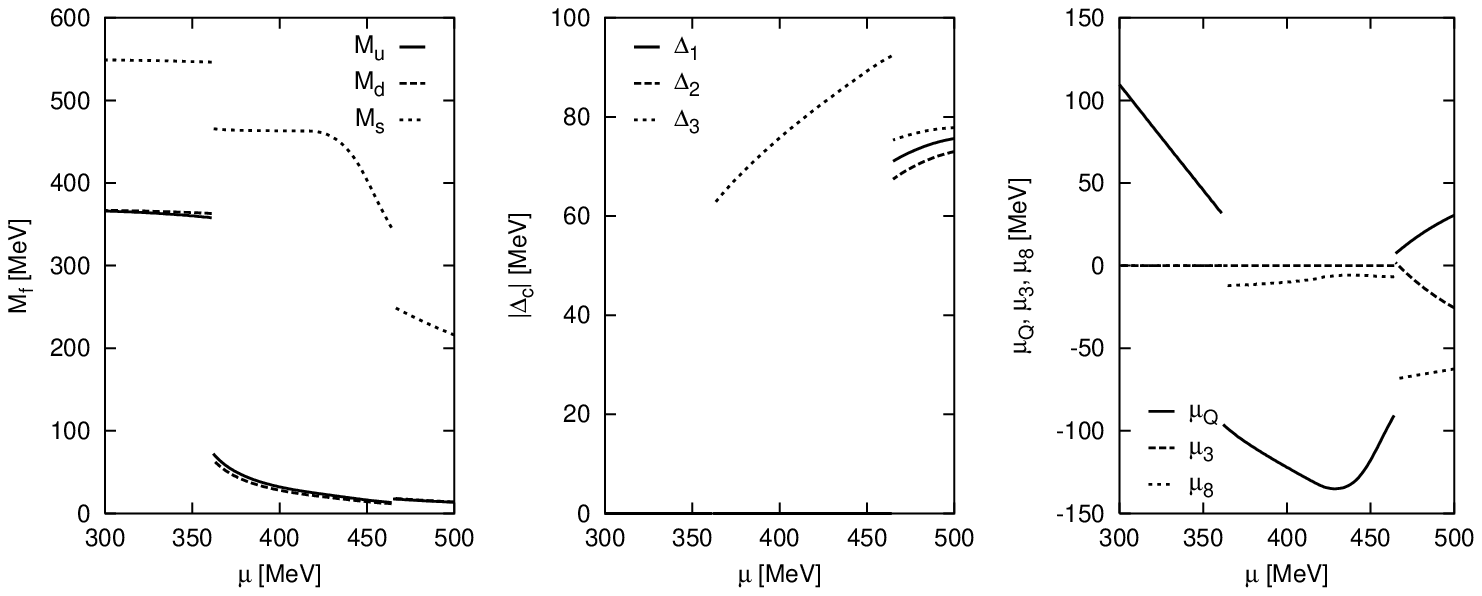}\\
    \includegraphics[width=0.95\textwidth]{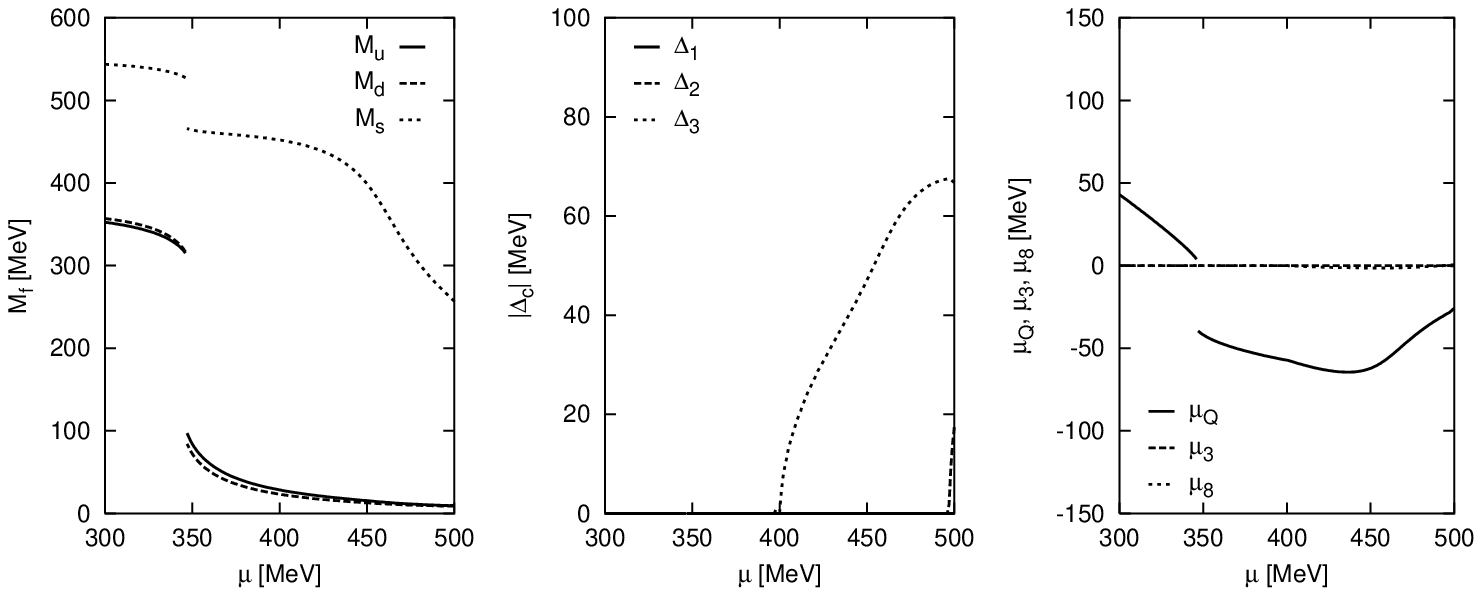}
    \caption{Dependence of the quark masses, of the gap parameters,
     and of the electric and color charge chemical potentials on the
     quark chemical potential at a fixed temperature, $T=0~\mbox{MeV}$ 
     (three upper panels) 
     and $T=40~\mbox{MeV}$ (three lower panels). The lepton-number 
     chemical potential is kept fixed at $\mu_{L_e}=200$~MeV.}
    \label{plot0-20-40_200}
  \end{center}
\end{figure}

\begin{figure}[ht]
  \begin{center}
    \includegraphics[width=0.95\textwidth]{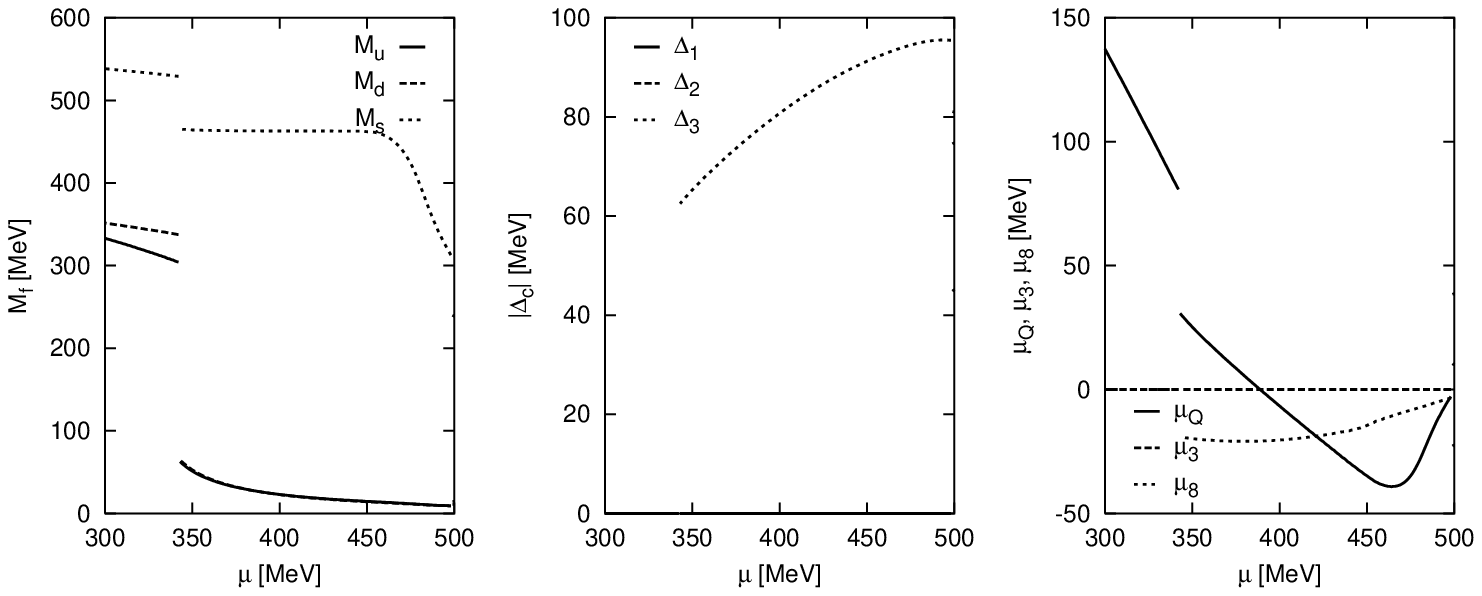}\\
    \includegraphics[width=0.95\textwidth]{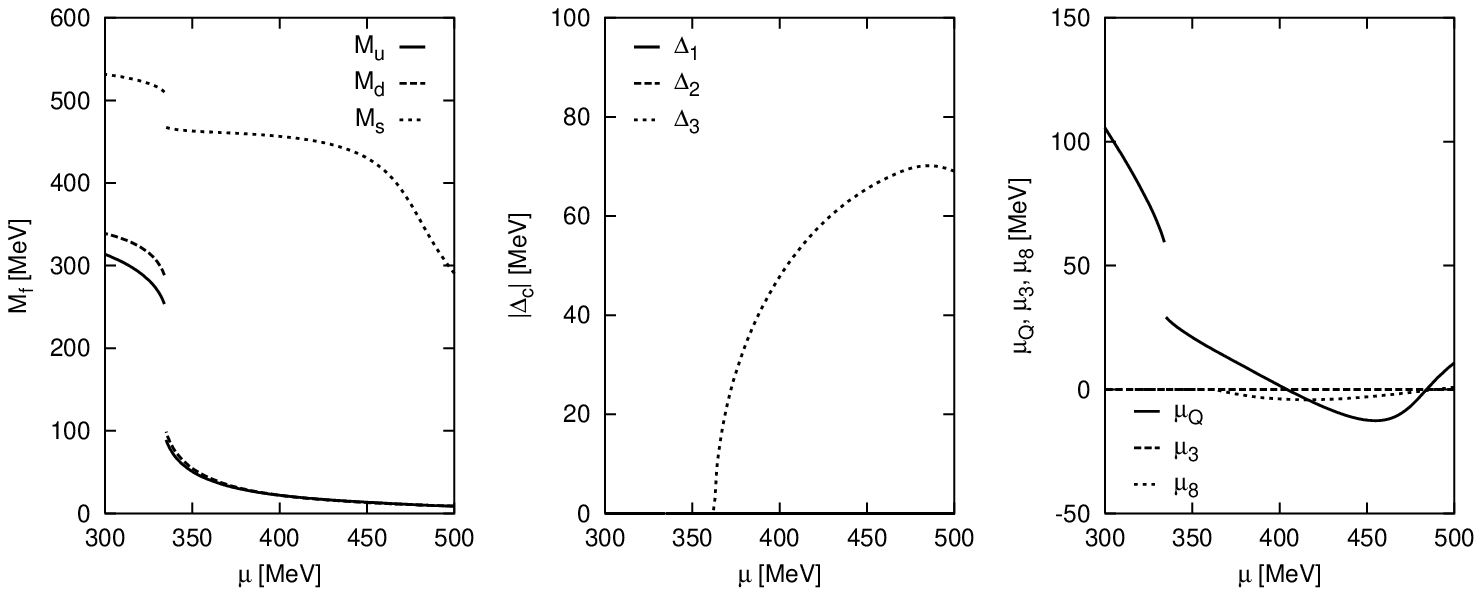}
    \caption{Dependence of the quark masses, of the gap parameters,
     and of the electric and color charge chemical potentials on the
     quark chemical potential at a fixed temperature, 
     $T=0~\mbox{MeV}$ (three upper panels), and 
     $T=40~\mbox{MeV}$ (three lower panels). The lepton-number 
     chemical potential is kept fixed at $\mu_{L_e}=400$~MeV.}
    \label{plot0-20-40_400}
  \end{center}
\end{figure}

In order to understand the basic characteristics of different phases 
in the phase diagrams in Fig.~\ref{phasediagram_mu-T}, we also present 
the results for the dynamical quark masses, the gap parameters, and the 
charge chemical potentials. These are plotted as functions of the quark 
chemical potential in Figs.~\ref{plot0-20-40_200} and \ref{plot0-20-40_400}, 
for two different values of the temperature in the case of 
$\mu_{L_e}=200$~MeV and $\mu_{L_e}=400$~MeV, respectively.

As in the absence of neutrino trapping, see Fig.~\ref{phasediagram}, 
there are roughly four distinct regimes in the phase diagrams in 
Fig.~\ref{phasediagram_mu-T}. At low temperature and low quark 
chemical potential, there is the $\chi$SB region. Here quarks 
have relatively large constituent masses which are close to the 
vacuum values, see Figs.~\ref{plot0-20-40_200} and 
\ref{plot0-20-40_400}. We note that the $\chi$SB phase is rather 
insensitive to the presence of a nonzero lepton-number chemical 
potential. With increasing $\mu_{L_e}$ the phase boundary is only 
slightly shifted to lower values of $\mu$. This is just another 
manifestation of the strengthening of the 2SC phase due to neutrino 
trapping.

With increasing temperature, the $\chi$SB phase turns into the 
normal quark matter phase, whose qualitative features are little 
affected by the lepton-number chemical potential. 

The third regime is located at relatively low temperatures 
but at quark chemical potentials higher than in the $\chi$SB 
phase. In this region, the masses of the up and down quarks 
have already dropped to values well below their respective
chemical potentials while the strange quark mass is still 
large, see left columns of panels in Figs.~\ref{plot0-20-40_200} 
and \ref{plot0-20-40_400}. As a consequence, up and down quarks 
are quite abundant but strange quarks are essentially absent.

It turns out that the detailed phase structure in this region is very 
sensitive to the lepton-number chemical potential. At $\mu_{L_e}=0$, 
the pairing between up and down quarks is strongly hampered 
by the constraints of neutrality and $\beta$ equilibrium, see 
Fig.~\ref{phasediagram}. The situation changes dramatically 
with increasing the value of the lepton-number chemical 
potential, see Fig.~\ref{phasediagram_mu-T}. The 
low-temperature region of the normal phase of quark matter 
is replaced by the (g)2SC phase (e.g., at $\mu = 400$~MeV, 
this happens at $\mu_{L_e} \simeq 110$~MeV). With $\mu_{L_e}$ 
increasing further, no qualitative changes happen in this part 
of the phase diagram, except that the area of the (g)2SC phase 
expands slightly.

Finally, the region in the phase diagram at low temperatures 
and large quark chemical potentials corresponds to phases 
in which the cross-flavor strange-nonstrange Cooper pairing 
becomes possible. In general, as the strength of pairing 
increases with the quark chemical potential, the system 
passes through regions of the gapless uSC (guSC), uSC, and 
gCFL phases and finally reaches the CFL phase. (Of course, the 
intermediate phases may not always be realized.) The effect 
of neutrino trapping, which grows with increasing the lepton-number 
chemical potential, is to push out the location of the 
strange-nonstrange pairing region to larger values of $\mu$. 
Of course, this is in agreement with the general arguments in 
Sec.~\ref{simple}.

As we mentioned earlier, the presence of the lepton-number chemical 
potential $\mu_{L_e}$ leads to a change of the quark Fermi 
momenta. This change in turn affects Cooper pairing of quarks, 
facilitating the appearance of some phases and suppressing 
others. As it turns out, there is also another qualitative 
effect due to a nonzero value of $\mu_{L_e}$. In particular, 
we find several new variants of gapless phases which did not 
exist at vanishing $\mu_{L_e}$. In Figs.~\ref{phasediagram_mu-T}
and \ref{phasediagramTnu}, these are denoted by the same names, 
g2SC or gCFL, but with one or two primes added. 

We define the g2SC$^{\prime}$ as the gapless two-flavor 
color-superconducting phase in which the gapless excitations 
correspond to $rd$--$gu$ and $gd$--$ru$ modes instead of the 
usual $ru$--$gd$ and $gu$--$rd$ ones, i.e., $u$ and $d$ 
flavors are exchanged as compared to the usual g2SC phase.
The g2SC$^{\prime}$ phase becomes possible only 
when the value of $(\mu_{ru} -\mu_{gd})/2\equiv(\mu_Q+\mu_3)/2$ 
is positive and larger than $\Delta_3$. The other phases are 
defined in a similar manner. In particular, the gCFL$^{\prime}$ phase,
already introduced in Sec.~\ref{simple}, is indicated by the gapless 
$rs$--$bu$ mode, while the gCFL$^{\prime\prime}$ phase has both 
$gs$--$bd$ (as in the gCFL phase) and $rs$--$bu$ gapless modes.
The definitions of all gapless phases are summarized in the following
table:
\begin{table}[h]
\begin{tabular}{|l|l|l|}
\hline
Name    & \begin{tabular}{l}Gapless mode(s)\\ 
        $\epsilon(k)\sim |k-k_F^{\rm eff}|$ \end{tabular} & Diquark condensate(s)\\
\hline
\hline
g2SC                  & $ru$--$gd$, $gu$--$rd$ & $\Delta_3$ \\
g2SC$^{\prime}$       & $rd$--$gu$, $gd$--$ru$ & $\Delta_3$ \\
guSC                  & $rs$--$bu$             & $\Delta_2$, $\Delta_3$  \\
gCFL                  & $gs$--$bd$             & $\Delta_1$, $\Delta_2$, $\Delta_3$ \\
gCFL$^{\prime}$       & $rs$--$bu$             & $\Delta_1$, $\Delta_2$, $\Delta_3$\\
gCFL$^{\prime\prime}$ & $gs$--$bd$, $rs$--$bu$ & $\Delta_1$, $\Delta_2$, $\Delta_3$\\
\hline
\end{tabular}
\end{table}

\subsection{Lepton fraction $Y_{L_e}$}

\begin{figure}[ht]
  \begin{center}
    \includegraphics[angle=-90,width=0.75\textwidth]{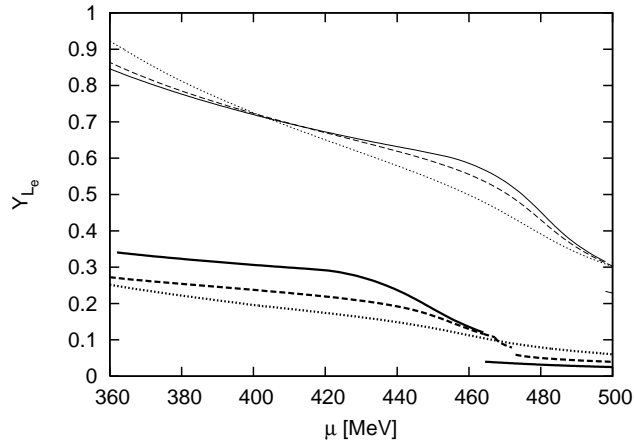}
    \caption{Dependence of the electron family lepton fraction $Y_{L_e}$
     for $\mu_{L_e} = 200$~MeV (thick lines) and $\mu_{L_e} = 400$~MeV
     (thin lines) on the quark chemical potential at a fixed temperature, 
     $T=0~\mbox{MeV}$ (solid lines), $T=20~\mbox{MeV}$ (dashed lines), and 
     $T=40~\mbox{MeV}$ (dotted lines).}
    \label{plot_Y_0-20-40}
  \end{center}
\end{figure}

Our numerical results for the lepton fraction $Y_{L_e}$ are shown 
in Fig.~\ref{plot_Y_0-20-40}. The thick and thin lines correspond
to two different fixed values of the lepton-number chemical potential,
$\mu_{L_e}=200$~MeV and $\mu_{L_e}=400$~MeV, respectively. For a fixed
value of $\mu_{L_e}$, we find that the lepton fraction changes only 
slightly with temperature. This is concluded from the comparison of 
the results at $T=0$ (solid lines), $T=20$~MeV (dashed lines), and 
$T=40$~MeV (dotted lines) in Fig.~\ref{plot_Y_0-20-40}.

As is easy to check, at $T=40$~MeV, i.e., when Cooper 
pairing is not so strong, the $\mu$ dependence of $Y_{L_e}$ 
does not differ very much from the prediction in the simple 
two-flavor model in Sec.~\ref{simple}. By saying this, of course, 
one should not undermine the fact that the lepton fraction in 
Fig.~\ref{plot_Y_0-20-40} has a visible structure in the 
dependence on $\mu$ at $T=0$ and $T=20$~MeV. This indicates 
that quark Cooper pairing plays a nontrivial role in determining 
the value of $Y_{L_e}$. 

Our numerical study shows that it is hard to achieve values 
of the lepton fraction more than about $0.05$ in the CFL 
phase. Gapless versions of the CFL phases, on the other 
hand, could accommodate a lepton fraction up to about 
$0.2$ or so, provided the quark and lepton-number chemical 
potentials are sufficiently large. 

{From} Fig.~\ref{plot_Y_0-20-40}, we can also see that the 
value of the lepton fraction $Y_{L_e} \approx 0.4$, i.e.,
the value expected at the center of the protoneutron star 
right after its creation, requires the lepton-number chemical 
potential $\mu_{L_e}$ in the range somewhere between 
$200$~MeV and $400$~MeV, or slightly higher. The larger
the quark chemical potential $\mu$ the larger $\mu_{L_e}$
is needed. Then, in a realistic construction of a star,  
this is likely to result in a noticeable gradient of
the lepton-number chemical potential at the initial time. 
This gradient may play an important role in the subsequent
deleptonization due to neutrino diffusion through dense 
matter.

\subsection{$T$--$\mu_{L_e}$ phase diagram}

\begin{figure}[ht]
  \begin{center}
    \includegraphics[width=0.75\textwidth]{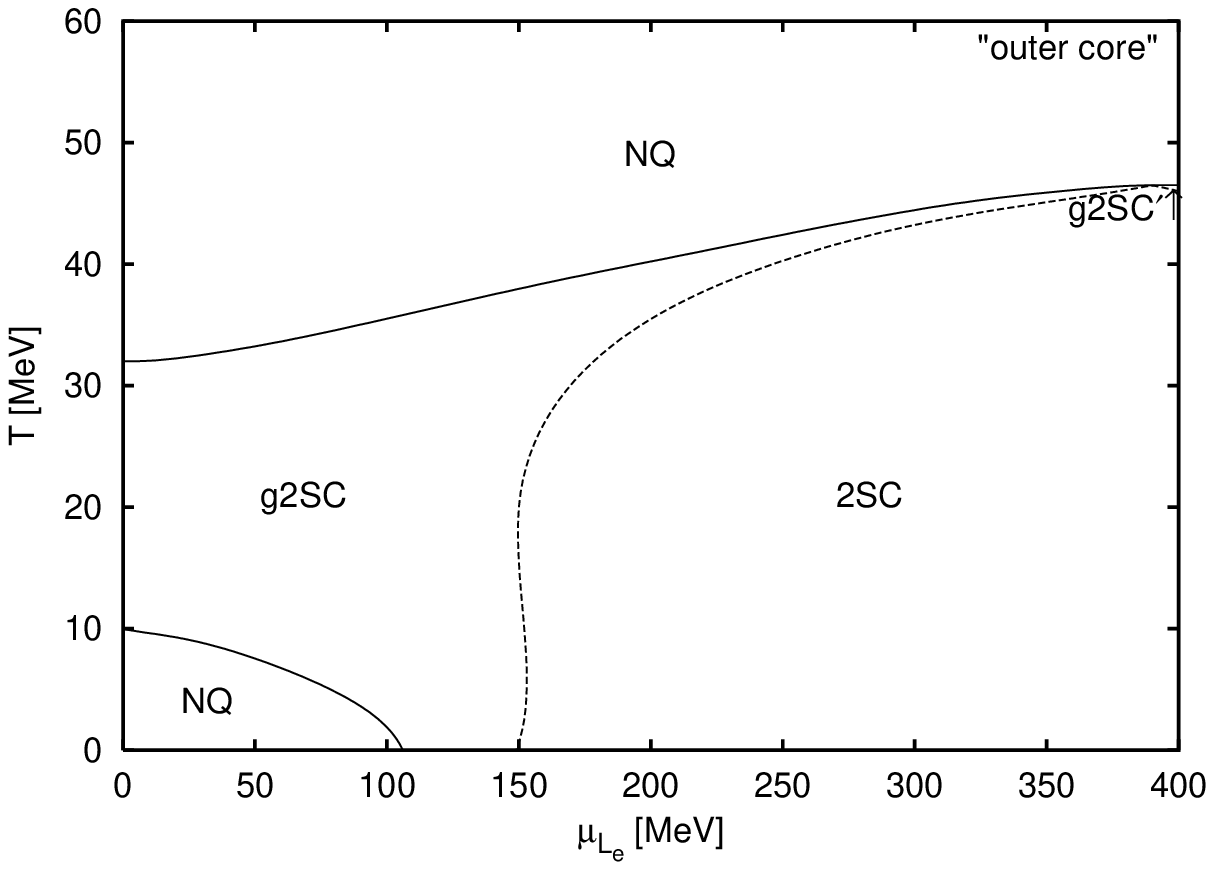}\\
    \includegraphics[width=0.75\textwidth]{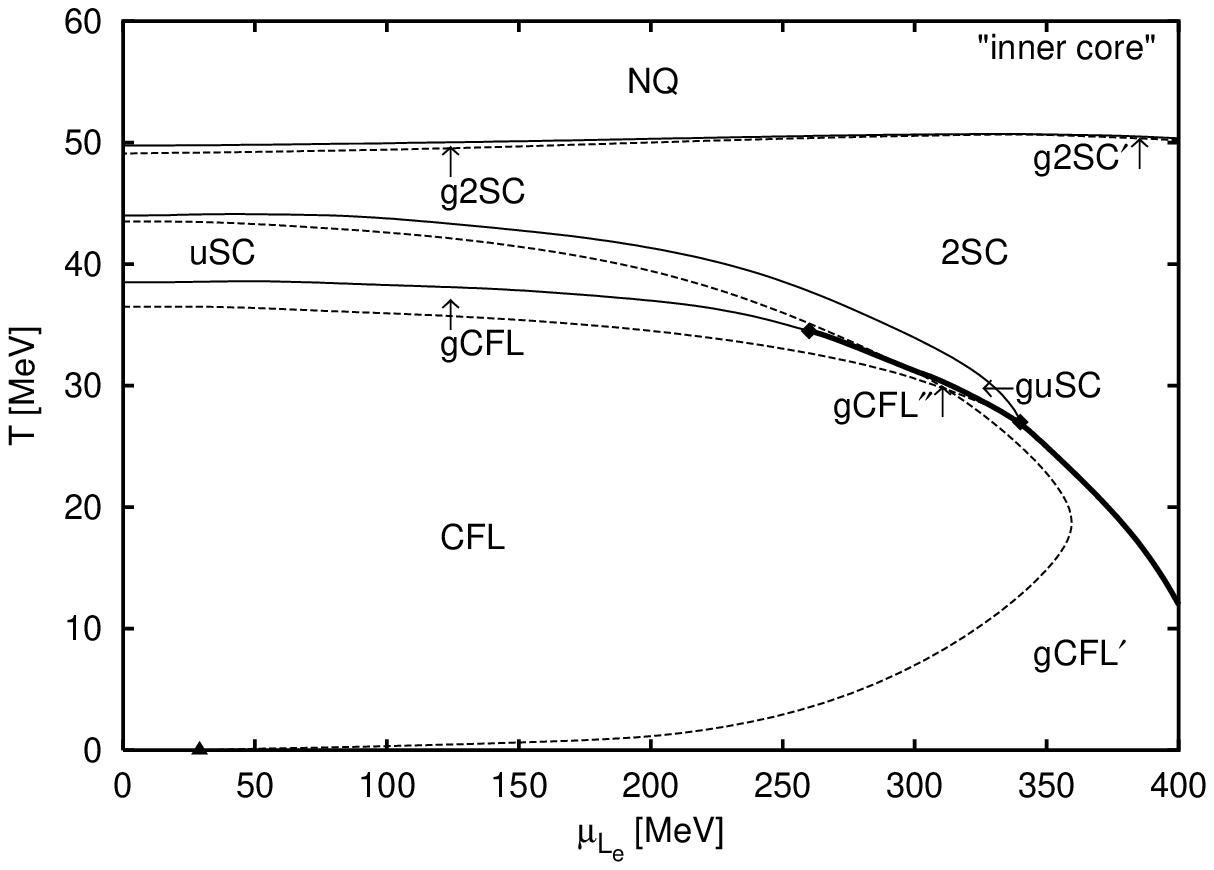}
    \caption{The phase diagrams of neutral quark matter in the plane of
      temperature and lepton-number chemical potential at two fixed
      values of quark chemical potential: $\mu=400$~MeV (upper panel)
      and $\mu=500$~MeV (lower panel). The triangle in the lower panel 
      denotes the transition point from the CFL phase to the
      gCFL$^{\prime}$ phase at $T=0$.}
  \label{phasediagramTnu}
  \end{center}
\end{figure}

Now let us discuss the phase diagram in the plane of temperature 
and lepton-number chemical potential, keeping the quark chemical 
potential fixed. Two such slices of the phase diagram are presented 
in Fig.~\ref{phasediagramTnu}. The upper panel corresponds to a 
moderate value of the quark chemical potential, $\mu=400$~MeV. 
This could be loosely termed as the ``outer core'' phase 
diagram. The lower panel in Fig.~\ref{phasediagramTnu} corresponds 
to $\mu=500$~MeV, and we could associate it with the ``inner 
core'' case. Note, however, that the terms ``inner core'' 
and ``outer core'' should not be interpreted literally here. The 
central densities of (proto-)neutron stars are subject to large 
theoretical uncertainties and, thus, are not known very well. 
In the model at hand, the case $\mu = 400$ MeV (``outer core") 
corresponds to a range of densities around $4\rho_0$, while the 
case $\mu = 500$ MeV (``inner core") corresponds to a range of 
densities around $10\rho_0$. These values are of the same order 
of magnitude that one typically obtains in models (see, e.g., 
Ref.~\refcite{prakash-et-al}).

At first sight, the two diagrams in Fig.~\ref{phasediagramTnu} 
look so different that no obvious connection between them could 
be made. It is natural to ask, therefore, how such a dramatic 
change could happen with increasing the value of the quark chemical 
potential from $\mu=400$~MeV to $\mu=500$~MeV. In order to understand 
this, it is useful to place the corresponding slices of the phase 
diagram in the three-dimensional diagram in Fig.~\ref{phase3d}. 

The $\mu=500$~MeV diagram corresponds to the right-hand-side 
surface of the bounding box in Fig.~\ref{phase3d}. This 
contains almost all complicated phases with strange-nonstrange 
cross-flavor pairing. The $\mu=400$~MeV diagram, on the other 
hand, is obtained by cutting the three-dimensional diagram 
with a plane parallel to the bounding surface, but going 
through the middle of the diagram. This part of the diagram 
is dominated by the 2SC and NQ phases. Keeping in mind the 
general structure of the three-dimensional phase diagram, 
it is also not difficult to understand how the two diagrams 
in Fig.~\ref{phasediagramTnu} transform into each other.

Several comments are in order regarding the zero-temperature
phase transition from the CFL to gCFL$^{\prime}$ phase, shown  
by a small black triangle in the phase diagram at $\mu=500$~MeV, 
see the lower panel in Fig.~\ref{phasediagramTnu}. The appearance 
of this transition is in agreement with the analytical result in 
Sec.~\ref{simple}. Moreover, the critical value of the lepton-number
chemical potential also turns out to be very close to the estimate
in Eq.~(\ref{mu_L^cr}). Indeed, by taking into account that 
$M_{s}\approx 214$~MeV and $\Delta_2\approx 76$~MeV, we obtain 
$\mu_{L_e}^{\rm (cr)}=\Delta_2-M_s^2/(2\mu) \approx 30$~MeV 
which agrees well with the numerical value. 

Before concluding this subsection, we should mention that a 
schematic version of the phase diagram in $T$--$\mu_{L_e}$ plane
was earlier presented in Ref.~\refcite{SRP}, see the left panel in
Fig.~4 there. In Ref.~\refcite{SRP}, the value of the quark 
chemical potential was $\mu=460$~MeV, and therefore a direct 
comparison is not straightforward. One can see, 
however, that the diagram of Ref.~\refcite{SRP} fits naturally 
into the three-dimensional diagram in Fig.~\ref{phase3d}. 
Also, the diagram of Ref.~\refcite{SRP} is topologically close
to the $\mu=500$~MeV phase diagram shown 
in the lower panel of Fig.~\ref{phasediagramTnu}. The 
quantitative difference is not surprising: the region of the 
(g)CFL phase is considerably larger at $\mu=500$~MeV than 
at $\mu=460$~MeV.

\section{Conclusions}
\label{conclusions}

Here we discussed the phase diagram of neutral three-flavor 
quark matter in the space of three parameters: temperature 
$T$, quark chemical potential $\mu$, and lepton-number 
chemical potential $\mu_{L_{e}}$. The analysis is performed in 
the mean-field approximation in the phenomenologically motivated 
NJL model introduced in Ref.~\refcite{RKH}. Constituent quark masses
are treated as dynamically generated quantities. The overall 
structure of the three-dimensional phase diagram is summarized 
in Fig.~\ref{phase3d} and further detailed in several 
two-dimensional slices, see Figs.~\ref{phasediagram}, 
\ref{phasediagram_mu-T} and \ref{phasediagramTnu}. 

By making use of simple model-independent arguments, as well as 
detailed numerical calculations in the framework of an NJL-type
model, we find that neutrino trapping helps Cooper pairing in
the 2SC phase and suppresses the CFL phase. In essence, this
is the consequence of satisfying the electric neutrality 
constraint in the quark system. In two-flavor quark matter, 
the (positive) lepton-number chemical potential $\mu_{L_e}$ 
helps to provide extra electrons without inducing a large 
mismatch between the Fermi momenta of up and down quarks. 
With reducing the mismatch, of course, Cooper pairing gets 
stronger. This is in sharp contrast to the situation in the
CFL phase of quark matter, which is neutral in the absence 
of electrons. Additional electrons due to large $\mu_{L_e}$ 
can only put extra stress on the system.

In application to protoneutron stars, the findings presented 
here suggest that the CFL phase is very unlikely to appear
during the early stage of the stellar evolution before the 
deleptonization is completed. If color superconductivity occurs
there, the 2SC phase is the best candidate for the ground state.
In view of this, it might be quite natural to suggest 
that matter inside protoneutron stars contains little or no 
strangeness (just as the cores of the progenitor stars) during 
the early times of their evolution. In this connection, it is 
appropriate to recall that neutrino trapping also suppresses 
the appearance of strangeness in the form of hyperonic matter 
and kaon condensation.\cite{prakash-et-al} The situation in 
quark matter, therefore, is a special case of a generic 
property.

After the deleptonization occurs, it is possible that the ground 
state of matter at high density in the central region of the star 
is the CFL phase. This phase contains a large number of strange
quarks. Therefore, an abundant production of strangeness should
happen right after the deleptonization in the protoneutron star. 
If realized in nature, in principle this scenario may have 
observational signatures.

\section*{Acknowledgments}
This review is based on work of Refs.~\refcite{pd-mass} and 
\refcite{pd-nu}. This work was supported in part by the Virtual 
Institute of the Helmholtz Association under grant No. VH-VI-041, 
by the Gesellschaft f\"{u}r Schwerionenforschung (GSI), and by 
the Deutsche Forschungsgemeinschaft (DFG).

\end{document}